\documentclass[12pt]{iopart}
\usepackage{epsfig}
\begin{document}

\jl{19}


\topical[Reaction models to probe the structure of light exotic nuclei]
{Reaction models to probe the structure of light exotic nuclei}

\author{Jim Al-Khalili$^1$ and Filomena Nunes$^2$}
\address{$^{1)}$School of Electronics and Physical Sciences,
University of Surrey, Guildford, GU2 7XH, U.K.\\
$^{2)}$National Superconducting Cyclotron Laboratory, Michigan State
University, East Lansing, Michigan 48824, USA }

\begin{abstract}
We review here theoretical models for describing various types of
reactions involving light nuclei on the driplines. Structure
features to be extracted from the analysis of such reaction data, as
well as those that need to be incorporated in the reaction models
for an adequate description of the processes, are also under focus.
The major unsolved theoretical issues are discussed, along with some
suggestions for future directions of the field.
\end{abstract}
\maketitle

\section{Introduction}

A number of experimental reviews on the physics of light exotic
nuclei \cite{tanrev,hansenrev,nigelrev} have focused mainly on
experimental techniques and the physics that could be extracted from
those measurements. Any attention paid to the reaction models used
in the analyses was rather modest, and has appeared mainly in
conference
proceedings \cite{isol,ian-rev,jakenam,jakhirsh}. Our aim in this
paper is to fill this gap by reviewing the progress in the theory of
modelling reactions with light dripline nuclei.

In the early days of our field, total reaction cross section
measurements were used to obtain information on halo nuclei. This
observable was one of the key pieces of evidence for the extended
density tails (large matter radii \cite{tan1,tan2}). However,
depending on the reaction model used, results could differ
significantly: if the appropriate granular structure of the
projectile was included in the reaction model \cite{alkha}, one
concluded, from the same data, that  halo nuclei were much larger
than predicted using one-body density models.

\begin{figure}
\begin{center}\includegraphics[%
	width=1.0\columnwidth]{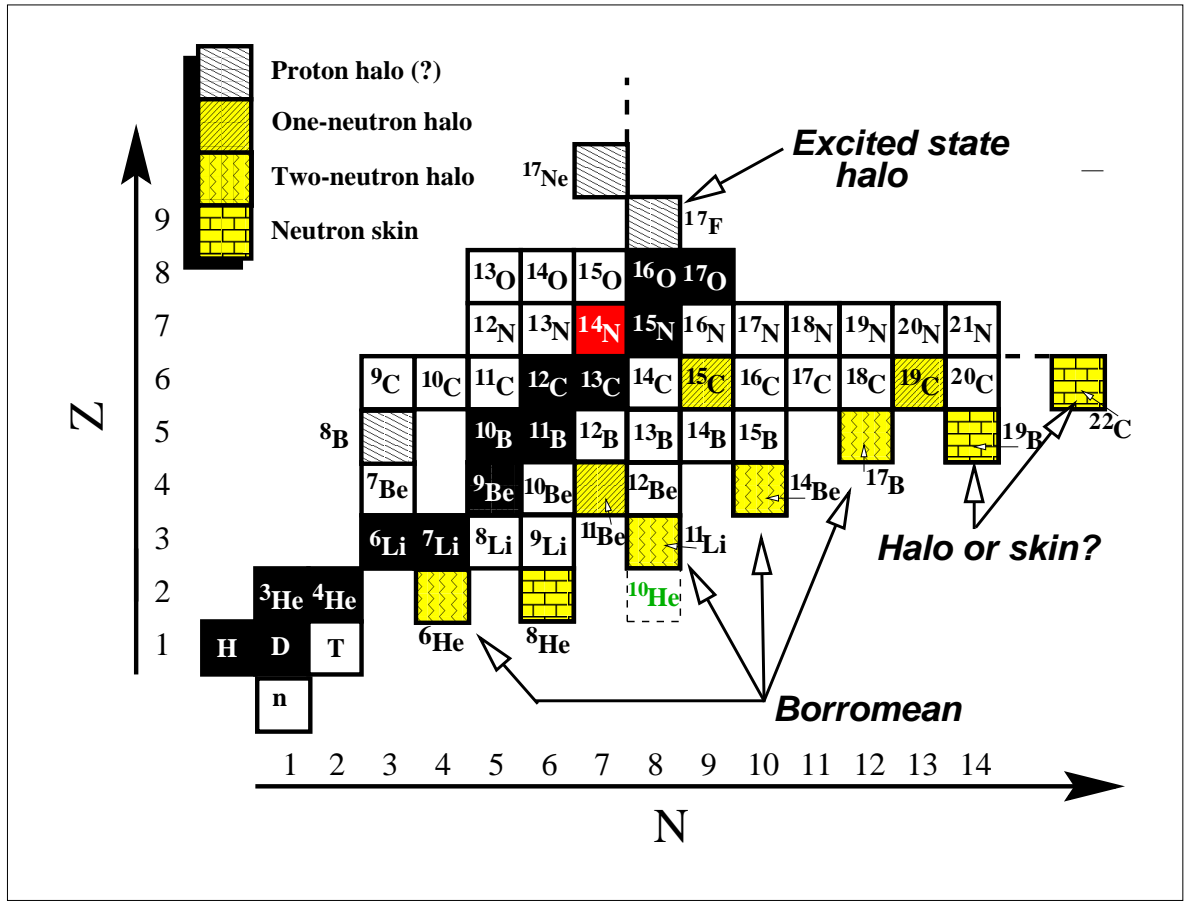}
\end{center}
\caption{\label{fig:segre}The light end of the chart of nuclides
showing where much of the current interest has been focused. Some of
the dripline nuclides found to exhibit new phenomena, such as halos,
are highlighted.}
\end{figure}

This is the first of many examples where there is an interplay
between structure and reaction. As the properties of these exotic
nuclei became evident (see figure (\ref{fig:segre})), reaction
models were modified in order to incorporate the essential known
structure features. The essence of these features could be
summarised as:
\begin{enumerate}

\item finite range effects extending out farther than expected,
due to the very long tails of the wavefunctions;

\item strong recoil effects due to the few-body granular structure;

\item continuum effects due to the proximity to threshold.
\end{enumerate}
The numerous cases discussed in this review are an illustration of
the importance of including these ingredients in the reaction model.

Following total reaction cross section data, and as soon as the beam
intensity allowed, elastic scattering for many of these nuclei was
measured, raising some paradigmatic questions about the type of
optical potentials required to fit the angular distributions that
are still not satisfactorily answered. However, it was the study of
breakup observables that attracted most of the theoretical effort.
Starting with momentum distributions, which essentially confirmed
the large spatial extension of the valence nucleons, progress lead
to experiments with complete kinematics, producing good quality
angular and energy distributions of the fragments.

Technical developments, both in the detection system and beam
production, enabled not only experiments with a larger variety of
exotic nuclei but, more importantly, measurements of the traditional
transfer and fusion reactions, the basis of most of the knowledge on
stable nuclei. Consequently, systematic measurements of knock-out
and transfer reactions, gave way to further theoretical
developments. The puzzling reports from recent fusion measurements
are presently a strong motivation for advances in the theory of
fusion reactions.

Most of the reaction theory for light nuclei on the driplines has
been developed for the high energy regime of fragmentation beams
where convenient approximations can be made. These include the
eikonal approximation, the adiabatic or {\it frozen halo}
approximation, first order perturbation theory, or even isolating
the nuclear and Coulomb transition amplitudes and treating them in
different ways. Fortunately, fragmentation data has been abundant,
providing crucial checks, allowing the identification of the exotic
features that need to be assimilated. Looking through the past two
decades, it is fair to say that significant progress has been made
and that it has been predominantly through the analysis of these high
energy data that we have learnt what we know about light dripline
nuclei.

In our view, a new era is beginning. We are no longer trying to
learn general features but the detailed structure. Unquestionably,
it is harder to model reactions at low energy. The  adiabatic and
eikonal simplifications are no longer expected to hold, couplings
are usually more important and isolating the nuclear and Coulomb
parts is often not possible. However, low energy data contain more
detailed information on the structure. The history of stable nuclear
physics shows that most of the detailed knowledge came from low
energy data. The large number of new generation ISOL facilities that
have or will become operational (e.g. SPIRAL-GANIL, REX-ISOLDE, EXCYT,
MAFF, ISAC-TRIUMF, HRIBF(Oak-Ridge), RIA-ANL, E-arena(JHF)) will
ensure progress in this direction.

In this review we report on both high energy and
low energy reactions models. Section 2 examines the theoretical
tools of various useful methods. Here we discuss the basic ideas
behind each method, summarising the formalism and providing the
relevant references. In Section 3, the models for analysis of total
reaction cross sections are presented. In section 4, elastic and
inelastic scattering studies are considered. Section 5
will cover the range of breakup models presently in use, from the
coupled channels CDCC method to the traditional DWBA, including time
dependent, semiclassical approaches as well as the models for high
energy reactions, mostly applied to momentum distributions.  In
Section 6, the model used to analyze knock-out data is discussed,
followed by a discussion of the applications of transfer reactions
to extract structure information (Section 7). In Section 8, we give
an account of the present status of fusion models. In section 9 we
look at the modeling of other types of
reactions which do not fit in any of the previous sections. Finally,
in Section 10, we conclude with a discussion on some of the main
open issues that need to be tackled in the near future, including
theoretical considerations for
reactions with electron beams.

\section{Theoretical tools}

In this section we give a brief outline of several approaches for
calculating reaction observables for light exotic nuclei. The common
feature of these systems is their weak binding and few-body nature.
It is often therefore important to treat their reactions within
few-body models also. We thus begin by discussing a number of
theoretical techniques which provide approximate descriptions of the
scattering and reactions of composite nuclei over a wide range of
incident energies.

\subsection{Few-body model space}
In general, we require approximate solutions of the time-independent
few-body Schr\"odinger equation. In this review, we focus mainly on
projectiles which, to a good approximation, can be described as
strongly-correlated $n$-body systems, where the $n$ constituents can
be individual nucleons or more massive clusters of many nucleons.
The projectile's ground state is assumed to be a bound state,
$\phi_0^{(n)}$ of the $n$ constituents, each of which can interact
with a target nucleus via complex two-body effective interaction,
$V_{jT}$. This potential is identified with the energy-dependent
phenomenological optical potential obtained by fitting reaction data
for the $j+T$ binary system at the same incident energy per nucleon
as the full projectile. If such data are not available then these
potentials are calculated, either from folding models or, more
microscopically, from multiple scattering theory.

In most cases of interest, the projectile nucleus has only one or
two particle stable bound states, which couple strongly to the
continuum during the reaction process. A major feature of
few-body reaction models is therefore the inclusion of such
projectile breakup effects in the reaction theory.

\begin{figure}
\begin{center}\includegraphics[%
	width=0.7\columnwidth]{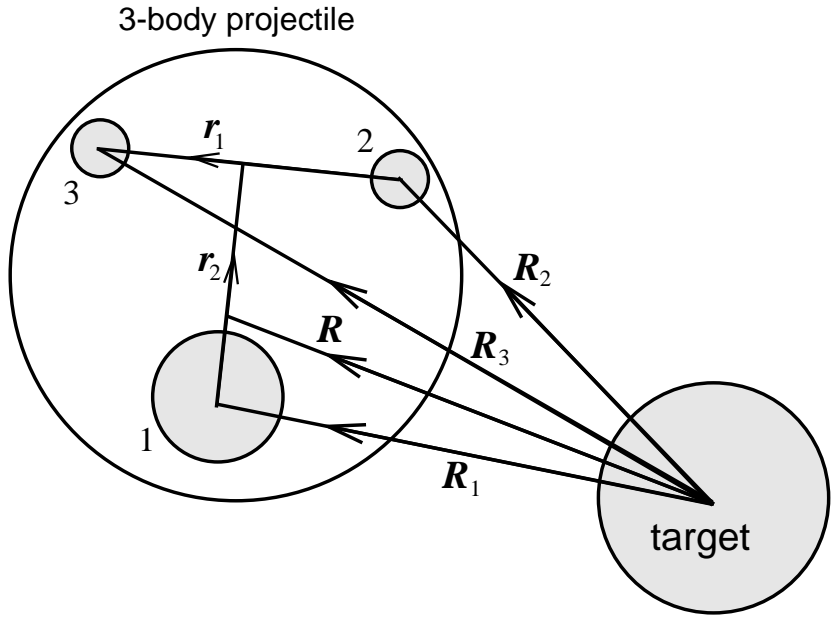}
\end{center}
\caption{\label{fig:cluster}Definition of the coordinate vectors
used in the case of the scattering of a three-body projectile from a
target.}
\end{figure}

The Schr\"{o}dinger equation satisfied by the scattering
wavefunction of the effective $(n+1)$-body (projectile and target)
system, $\Psi^{(+)}$, when the projectile is incident with wave
vector $\vec{K}_0$ in the cm frame, is
\begin{equation}
\left[\, H\,-\,E \right] \Psi_{\vec{K}_0}^{(+)} (\vec{R},\vec{r}_1,\cdots,
\vec{r}_n)=0~~, \label{seqn}
\end{equation}
with total Hamiltonian
$H=T_{R}+U(\vec{R}_1,\cdots,\vec{R}_n)+H_{p}$. Here $H_{p}$ is the
internal Hamiltonian for the projectile and $T_{R}$ is the
projectile cm kinetic energy operator. The vectors $\{\vec{r}_i\}$
are the relative (internal) coordinates between the projectile
constituents, and $\{\vec{R}_j \}$ are the position vectors of the
projectile constituents with respect to the target (see figure
(\ref{fig:cluster})). The total interaction between the projectile
and target is just the sum of projectile constituent-target
interactions:
\begin{equation}
U(\vec{R}_1,\cdots,\vec{R}_n) =\sum_{j=1}^n V_{jT}(R_j)\ .\label{potsum}
\end{equation}
The $n$-body projectile ground state wavefunction $\phi^{(n)}_0$ satisfies
\begin{equation}
H_{p} \,\phi^{(n)}_0 (\vec{r}_1,\cdots,\vec{r}_n) = -\varepsilon_0 \,
\phi^{(n)}_0 (\vec{r}_1,\cdots,\vec{r}_n)~.
\end{equation}
$H_p$ will also generate an excited continuum spectrum and may
support a finite number of bound or resonant excited states. We thus
seek solutions of the few-body scattering wave function
$\Psi_{\vec{K}_0 }^{(+)}$ which satisfy the scattering boundary
conditions
\begin{equation}
\Psi_{\vec{K}_0}^{(+)}=e^{i\vec{K}_0\cdot
\vec{R}}\,\phi^{(n)}_0 \,+ \,{\rm outgoing\ waves}~~,
\label{bcns}
\end{equation}
and where the target nucleus is assumed to remain in its ground
state. For a projectile with a single bound state, the outgoing
waves include only elastic scattering and elastic break-up channels.
More generally, the outgoing waves will also include terms from any
inelastically excited bound states.

It is implicit in the following that the methods we discuss yield
only approximate solutions of the physical $n$-body problem. In
particular one- and multi-constituent rearrangement channels are
absent in the asymptotic ($R \rightarrow \infty$) regions of the
derived solutions, due to our use of complex constituent-target
interactions and radial and orbital angular momentum
truncations\cite{PhysRep}. In fact, all the theoretical schemes
calculate approximations to $\Psi^{(+)}$ which are expected to be
accurate representations of the $n$-body dynamics only within a
restricted volume of the configuration space, or within a given
interaction region. Reaction or scattering amplitudes can
nevertheless be calculated reliably by using the wavefunction within
an appropriate transition matrix element.

\subsection{Continuum discretisation method}
\label{cdcc}

The most accurate theoretical technique available for reactions
involving a projectile that can be reliably modeled as a two-cluster
system is the method of coupled discretised continuum channels
(CDCC) \cite{PhysRep}. It was originally formulated and applied to
the scattering of the deuteron (n+p), $^6$Li ($\alpha$+d) and $^7$Li
($\alpha$+t), but has more recently been applied to a number of
loosely-bound core+valence nucleon modeled dripline nuclei. The
method cannot, however, be extended readily to three-body
projectiles, such as the Borromean nuclei ($^6$He and $^{11}$Li),
although progress in developing such a four-body CDCC model is being
made.

The CDCC method approximates the three-body Schr\"{o}dinger equation
as a set of effective two-body coupled-channel equations by
constructing a square integrable basis set $\{\phi_\alpha\}$ of
relative motion states between the two constituents of the
projectile (including as well as the bound states, a representation
of the continuum).

Projectiles treated using the CDCC method tend to have very few
(often just one) bound states and the method provides a means of
describing excitations to the continuum. Each of the physically
significant set of spin-parity relative motion excitations is
divided (or `binned') into a discrete set of energy or momentum
intervals up to some maximum value.

The CDCC method therefore works with the model space Hamiltonian
\begin{equation}
H^{CDCC} = PHP,\hspace{2em} P=\sum_{\alpha=0}^N|\phi _{\alpha}
\rangle \langle \phi_{\alpha}|,
\end{equation}
where the subscript $\alpha$ refers to the set of discrete states
(ground state plus excited states) corresponding to energy
eigenvalues $\varepsilon_\alpha = \langle \phi_\alpha |H_p
|\phi_\alpha \rangle$. The corresponding asymptotic wavenumbers
$K_\alpha$, associated with the cm motion of the projectile in these
excited configurations, are such that
\begin{eqnarray}
{\hbar^2 K_\alpha^2}/{2\mu_{p}}+\varepsilon_\alpha={\hbar^2 K_0^2}
/{2\mu_{p}} - \varepsilon_0=E~.
\end{eqnarray}

These bin states, together with the ground state, constitute an
$(N+1)$ state coupled-channels problem for
solution of the CDCC approximation to $\Psi^{(+)}$
\begin{eqnarray}
\Psi_{\vec{K}_0}^{CDCC}(\vec{r},\vec{R}) = \sum_{\alpha=0}^{N}
\phi_{\alpha}(\vec{r}) \chi_{\alpha}(\vec{R}),
\end{eqnarray}
where $\alpha=0$ refers to the projectile ground state. Explicitly
\begin{eqnarray}
\left[\,T_{R} \,+\,V_{\alpha \alpha}(\vec{R})\,-\,E_\alpha \right]
\chi_{\alpha}(\vec{R})= -\sum_{\beta\neq\alpha}V_{\alpha\beta}
(\vec{R}) \chi_{\beta}(\vec{R}),
\end{eqnarray}
with $E_\alpha=E-\varepsilon_\alpha$. The coupling interactions are
\begin{eqnarray}
V_{\alpha \beta}(\vec{R}) = \langle \phi_{\alpha}
|U(\vec{R}_1,\vec{R}_2)| \phi_{\beta}\rangle ,
\end{eqnarray}
keeping in mind the definition of $\vec{R}$, $\vec{R}_1$ and
$\vec{R}_2$ as illustrated in figure (\ref{fig:cluster}).

The evaluation of these couplings involves additional practical
truncations of the CDCC model space, namely of $a)$ the maximum
order used in the multipole expansion of the interactions $U$, and
$b)$ the maximum radius $r_{bin}$ used in evaluating these matrix
elements. These must be chosen to be consistent with the included
$J_p^*$ channels, the bin widths $\Delta k_i$ and interaction
ranges.

Convergence of the calculations are then tested for different sizes
of this model space. The number of bins and their upper limit depend
on the particular state they are describing. The parameters must of
course be carefully chosen to map any characteristic or resonant
features in the projectile continuum. The different schemes for
construction of the bin states are discussed extensively in the
literature \cite{scatt}.

Solution of the coupled equations is carried out by usual partial
wave decomposition. This allows the calculation of the elastic or
inelastic scattering amplitude required for observables such as the
differential cross section angular distribution. The CDCC scheme is
available in a general coupled-channels computer code
\cite{thom1,thom2}.

Nuclear and Coulomb breakup of two-body projectiles, such as $^8$B,
$^{11}$Be, $^{17}$F and $^{19}$C, can also be calculated with this
model. The breakup transition amplitudes $T_{m} (\vec{k},\vec{K})$
from an initial state $J,m$ to a general physical three-body final
state of the constituents, with final state cm wave vector $\vec{K}$
and relative motion wave vector $\vec{k}$, is done by replacing
$\Psi^{CDCC}$ in an exact post-form matrix element \cite{kin3b},
\begin{eqnarray}
T_{m}(\vec{k},\vec{K})=\langle \phi_{\vec{k}}^{(-)}(\vec{r})
\,e^{i\vec{K} \cdot\vec{R}}|U|
\Psi_{\vec{K}_0 m}^{CDCC}(\vec{r},\vec{R})\rangle~~.
\end{eqnarray}
Inserting the set of bin-states, assumed complete for the model
space used, then allows us to write the transition amplitude as a
sum of amplitudes for each bin state, calculated by solving the
coupled equations \cite{scatt}.

\subsection{Adiabatic (sudden) approximation}
\label{adiamodel}

For reactions involving incident projectile energies above a few
tens of MeV per nucleon, a considerable simplification to the CDCC
method can be applied if we make use of an adiabatic treatment of
the dynamics. By identifying the energetic (fast) variable with the
projectile's cm motion coordinate, $\vec{R}$, and the slow variable
with the projectile's internal coordinates, $\vec{r}_i$, the
few-body Schr\"{o}dinger equation can be reduced to a much simpler
two-body form where the dynamical variable is only $\vec{R}$ and the
projectile's internal degrees of freedom enter only as parameters
(to be integrated over later). In the model, as formulated by
Johnson and Soper \cite{adia}, the approximation amounts to assuming
that the breakup energies $\varepsilon_k$ associated with the most
strongly coupled excitation configurations in equation (\ref{seqn})
are such that $\varepsilon_k\ll E$. Equivalently, due to the slow
internal motions of the constituents of the projectile, the
$\{\vec{r}_i\}$ are assumed frozen for the time taken for the
projectile to traverse the interaction region. This approximation is
also the starting point for the few-body Glauber method, based on
impact parameter descriptions, discussed in the next subsection.

The crucial step is to replace $H_{p}$ by $-\varepsilon_0$, the
projectile ground state binding energy. This is done to satisfy the
incident channel boundary conditions (the projectile is incident in
its ground state). What has been assumed here is that, while the
projectile does couple to excited and breakup states, they are all
taken to be degenerate with the energy of the dominant elastic
channel, $\varepsilon_0$. The adiabatic Schr\"odinger equation is
therefore, with $E_0=E+\varepsilon_0$,
\begin{equation}
\left[\,T_{R} \,+\,U\,-\,E_0\right]
\Psi_{\vec{K}_0}^{AD}(\vec{R},\vec{r}_1,\cdots,\vec{r}_n)=0~~.
\label{adeqn}
\end{equation}
The crucial point here is that the Hamiltonian now only has
parametric dependence on the projectile coordinates $\{\vec{r}_i\}$,
which appear in the potential, $U$.

Clearly, for two-body projectiles, the full CDCC approach is more
accurate than the adiabatic approach, particularly at low energies.
However, the adiabatic model does not suffer so much from
convergence issues or computational limitations. Also, it has been
generalised and applied to three-body projectiles, something the
CDCC method cannot yet cope with. In addition, the adiabatic model
allows for certain simplifying insights, such as when only one of
the projectile's constituents interact with the target. An example
of this is the pure Coulomb breakup of a one-neutron halo nucleus
like $^{11}$Be, to be discussed later on.

\subsection{Glauber methods}
\label{glauber}

A far more efficient approach for dealing with an $n$-body
projectile is to use the few-body Glauber (FBG) model, which is
based on the eikonal approximation.

The eikonal approximation was introduced in quantum scattering
theory by Moliere and later developed by Glauber who applied it to
nuclear scattering where he formulated a many-body, multiple
scattering generalisation of the method\cite{Glaub}. In common with
other semi-classical approaches, the eikonal method is useful when
the wavelength of the incident particle is short compared with the
distance over which the potential varies appreciably. This {short
wavelength condition} is expressed in terms of the incident centre
of mass wave number, $K_0$, and the range of the interaction, $R_0$,
such that
\begin{eqnarray}
K_0R_0 \gg 1\ .\label{conds1}
\end{eqnarray}
However, unlike short wavelength methods such as the WKB
approximation, the eikonal approximation also requires high
scattering energies, such that
\begin{eqnarray}
E \gg \vert V_0\vert \ ,
\end{eqnarray}
where $V_0$ is a measure of the strength of the potential. In
practice, and when $V$ is complex, this high energy condition is not
critical and the eikonal approximation works well even when
$E\approx \vert V_0\vert$ provided the first condition, equation
(\ref{conds1}), holds and we restrict ourselves to forward angle
scattering.

The eikonal wavefunction has incorrect asymptotics and so, to
calculate amplitudes and observables, it must be used within a
transition amplitude.  Thus for two-body elastic scattering, via a
central potential $V(R)$, the transition amplitude is
\begin{eqnarray}
T(\vec{K}_0,\vec{K}) =  \langle \vec{K} \vert~ V~\vert
{\psi}_{\vec{K}_0}^{eik} \rangle \ .
\end{eqnarray}
This leads to the well-known form of the scattering amplitude
\begin{eqnarray}
f_0(\theta)=-iK_0\!\int_{0}^{\infty}\!\! b\,{
d}b\,J_0(qb)
\left[S_0(b)\,-\,1\,\right]\ ,
\end{eqnarray}
where $q=2K_0\sin (\theta/2)$, $\theta$ is the scattering angle,
and $S_0(b) = \exp{[i\chi(b)]}$ is the
eikonal elastic $S$-matrix element at impact parameter $b$, and
the eikonal phase shift function, $\chi(b)$, is
\begin{eqnarray}
\chi(b)
=-\frac{1}{\hbar v}\int_{-\infty}^{\infty}\!\!\! V(R)\ dz\ .
\end{eqnarray}

\begin{figure}
\begin{center}\includegraphics[%
	width=0.7\columnwidth]{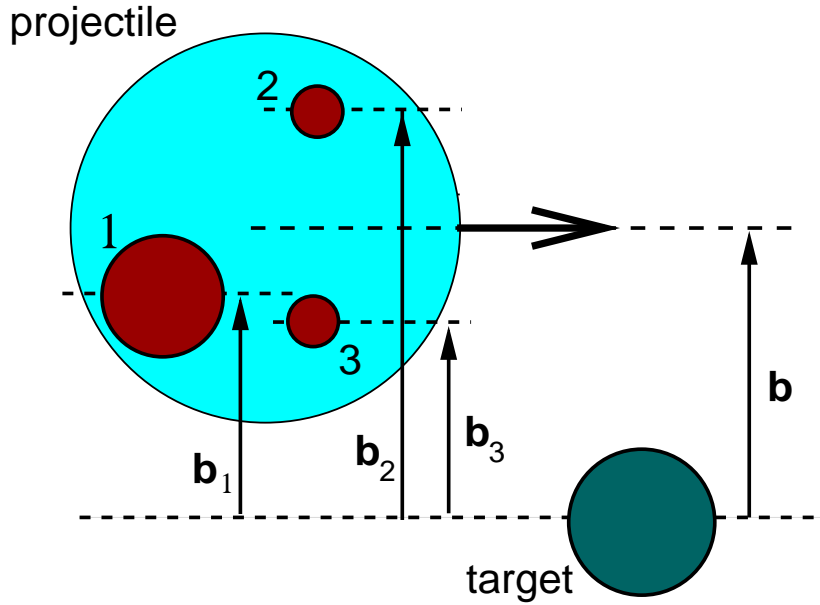}
\end{center}
\caption{\label{fig:eikon}In the few-body Glauber model, each
constituent of the projectile scatters from the target independently
with its $S$ matrix defined as a function of its own impact
parameter.}
\end{figure}

The few-body Glauber (FBG) scattering amplitude, for a collision
that takes a composite $n$-body projectile from an initial state
$\phi_0^{(n)}$ to a final state $\phi_\alpha^{(n)}$, can be derived
following the same steps as in the two-body (point particle
projectile) case. The post form transition amplitude is
\begin{eqnarray}
T(\vec{K}_\alpha)=\langle \phi^{(n)}_\alpha
\,e^{i\vec{K}_\alpha\cdot\vec{R}}\vert
U(\{\vec{R}_j\})\vert\Psi^{eik}_{\vec{K}_0}\rangle ,
\end{eqnarray}
and we obtain
\begin{eqnarray}
f^{(n)}(\vec{K}_\alpha) = -\frac{iK_0}{2\pi}\!\int\! d\vec{b}\,
e^{i\vec{q}\cdot\vec{b}}\,\langle\phi_\alpha^{(n)}\vert S^{(n)}
(\vec{b}_1,\cdots,\vec{b}_n)-1\vert\phi_0^{(n)}\rangle ,
\label{fbampnew}
\end{eqnarray}
where
\begin{eqnarray}
S^{(n)}=
\exp\left[i\sum_{j=1}^n \chi_j(b_j)\right]=\prod_{j=1}^{n}S_j(b_j)
.\label{fbs}
\end{eqnarray}
Thus the total phase shift is the sum of the phase shifts for the
scattering of each of the projectile's constituents, as shown in
figure (\ref{fig:eikon}). This property of phase shift additivity is
a direct consequence of the linear dependence of eikonal phases on
the interaction potentials $V_{jT}$.

Corrections to the straight line assumption of the eikonal
approximation have been calculated and allow the FBG approach to be
applied at much lower energies than expected (below 20 MeV/A). The
most straightforward approach is to replace the eikonal S-matrices
by the physical ones, while retaining the simplicity of the impact
parameter framework of the model \cite{brooke}.

The model generalises in a natural way when Coulomb forces are
included in the projectile constituent-target potentials, $V_{jT}$.

\subsection{The Optical Limit of the Glauber model}

The Glauber model can be simplified considerably at high energies
when the interaction between each projectile constituent and the
target can be considered as purely absorptive. In this case, each
constituent S-matrix, $S_j(b_j)$, is calculated within the optical
limit of the Glauber model \cite{alkhazov74}. Here, the eikonal
phase shifts are calculated assuming a '$t\rho\rho$' approximation
to the optical potentials, $V_{jT}$, using one-body densities for
each $j$ constituent and the target, and an effective
nucleon-nucleon amplitude, $f_{NN}$. The optical limit $S$-matrices
are thus written as
\begin{equation}
S_j^{OL}(b)= exp\left[i\int_{-\infty}^{\infty} dz\,\int\int
d\vec{r}_1\,d\vec{r}_2\,\rho_j(r_1)\rho_T(r_2)f_{NN}(\vert
\vec{R}+\vec{r}_1- \vec{r}_2 \vert)\right] .
\end{equation}
For an absorptive zero range NN amplitude and an isospin zero target
we have
\begin{equation}
f_{NN}(\vec{r})=(i\bar{\sigma}_{NN}/2)\delta(\vec{r})
\end{equation}
where $\bar{\sigma}_{NN}$ is the average of the free nn and np total
cross sections at the energy of interest.

It is important to note that we have not thrown away here the
few-body correlations in the projectile since at this stage it is
only the constituents' scattering via their individual $S_j$'s that
has been treated in the optical limit (OL). The few-body $S$-matrix
is still defined according to equation (\ref{fbs}). However, if all
few-body correlations are also neglected then $S^{(n)}$ is replaced
by $S^{OL}$ which is defined as for the individual $S^{OL}_j$ but
with $\rho_j$ replaced by the one-body density for the whole
projectile. In this case it can easily be shown that full
projectile-target OL $S$-matrix is equivalent to neglecting breakup
effects in equation (\ref{fbampnew}), i.e.,
\begin{eqnarray}
S^{OL}(b)=
\exp\left[\langle\phi_0^{(n)}\vert i\sum_{j=1}^n \chi_j(b_j)\vert
\phi_0^{(n)}\rangle \right] .
\end{eqnarray}

\subsection{Cross sections in Glauber theory}

The Glauber model provides a convenient framework for calculating
integrated cross sections for a variety of processes involving
peripheral collisions between composite projectiles and stable
targets. In particular, stripping reactions have been studied using
approaches developed by Serber\cite{Serber}. Variants of such
methods are still in use today due to the simple geometric
properties of the reaction processes at high energies.

In the few-body Glauber model, the differential cross section for the
scattering process defined by equation(\ref{fbampnew}) is
\begin{eqnarray}
\left(\frac{d\sigma}{d\Omega}\right)_{\alpha} = \vert f^{(n)}
(\vec{K}_\alpha)\vert^2 ,
\end{eqnarray}
and the total cross section for populating the final state $\alpha$
is thus
\begin{eqnarray}
\sigma_{\alpha}&=&\int d\Omega\ \vert f^{(n)}(\vec{K}_\alpha)\vert^2
\nonumber\\ &=&\int d\vec{b}\ \vert\langle
\phi^{(n)}_\alpha\vert S^{(n)}\vert\phi^{(n)}_0\rangle-
\delta_{\alpha 0} \vert^2\ .
\end{eqnarray}
It should again be noted however that such an expression is only
valid at high beam energies and low excitation energies since energy
conservation is not respected in this model. When $\alpha=0$, the
total elastic cross section is
\begin{eqnarray}
\sigma_{el}=\int d\vec{b}\,\,\vert 1 - \langle\phi^{(n)}_0\vert
S^{(n)}\vert\phi^{(n)}_0 \rangle \vert ^2 .\label{elsig}
\end{eqnarray}
The total cross section is also obtained from the elastic scattering
amplitude, employing the optical theorem, to give
\begin{eqnarray}
\sigma_{tot}=2\int d\vec{b} \left[ 1-\Re\,\langle\phi^{(n)}_0\vert
S^{(n)}\vert\phi^{(n)}_0 \rangle\right].\label{totsig}
\end{eqnarray}
Hence, the total reaction cross section, defined as the difference
between these total and elastic cross sections, is
\begin{eqnarray}
\sigma_{R}=\int d\vec{b} \left[ 1-\vert \langle\phi^{(n)}_0\vert
S^{(n)}\vert\phi^{(n)}_0 \rangle\vert^2\right].\label{reacsig}
\end{eqnarray}

For a projectile of total angular momentum $j$, the above
expression is more correctly:
\begin{eqnarray}
\sigma_{R}=\frac{1}{2j+1}\int d\vec{b} \sum_{m,m'}
\left[ 1-\vert \langle\phi^{(n)}_{0m'}\vert
S^{(n)}\vert\phi^{(n)}_{0m} \rangle\vert^2\right].\label{reacsig2}
\end{eqnarray}

For projectiles with just one bound state, any excitation due to
interaction with the target will be into the continuum. For such
nuclei, which include the deuteron and many of the neutron halo
nuclei (such as $^6$He and $^{11}$Li), it is possible to describe
elastic breakup channels in which the target as well as each cluster
in the projectile remain in their ground states. For simplicity of
notation, we assume a two-body projectile with continuum wave
functions $\phi_{\vec{k}}$, where $\vec{k}$ is the relative momentum
between the two clusters, and $S=S^{(2)}(b_1,b_2)=S_1(b_1)S_2(b_2)$
is understood. Elastic breakup, also referred to as diffractive
dissociation, has amplitudes
\begin{eqnarray}
f(\vec{k},\theta)= -iK_0\!\int d\vec{b}\,
e^{i\vec{q}\cdot\vec{b}}\,\langle\phi_{\vec{k}\sigma}\vert
S\vert\phi_{0m}\rangle.\label{fbgbu}
\end{eqnarray}
Making use of the completeness relation (when there is
only one bound state)
\begin{eqnarray}
\int d\vec{k} \,\,\vert\phi_{\vec{k}\sigma}\rangle
\langle\phi_{\vec{k}\sigma}\vert = 1-\vert\phi_{0m}\rangle
\langle\phi_{0m}\vert
\end{eqnarray}
the total elastic breakup cross section is
\begin{eqnarray}
\sigma_{bu}=\frac{1}{2j+1}\int d\vec{b} \sum_{m,m'}\left[
\langle\phi_{0m} \vert\ \vert S\vert^2
\vert\phi_{0m}\rangle \delta_{m,m'} - \vert \langle\phi_{0m'}\vert
S\vert\phi_{0m}\rangle\vert^2\right]\ .
\end{eqnarray}
The difference between the reaction and elastic breakup cross section
is the absorption cross section,
\begin{eqnarray}
\sigma_{abs}=\frac{1}{2j+1}\int d\vec{b} \sum_m \left[ 1-
\langle\phi_{0m}\vert\ \vert
S\vert^2\vert\phi_{0m} \rangle\right]\ ,
\end{eqnarray}
which represents the cross section for excitation of either the
target or one or both of the projectile clusters.

The above formula can be understood by examining the physical
meaning of $\vert S\vert^2$ ($=\vert S_1\vert^2\vert S_2\vert^2$).
The square modulus of each cluster $S$-matrix element, $\vert
S_i\vert^2$ represents the probability that it survives intact
following its interaction with the target at impact parameter
$\vec{b}_i$. That is, at most, it is elastically scattered. At large
impact parameters $\vert S_i\vert^2\rightarrow 1$ since the
constituent passes too far from the target.  The quantity $1-\vert
S_i\vert^2$ is therefore the probability that cluster $i$ interacts
with the target and is absorbed from the system. Such a simple
picture is useful when studying stripping reactions in which one or
more of the projectile's clusters are  removed by the target while
the rest of the projectile survives. Thus, the cross section for
stripping cluster $1$ from the projectile, with cluster $2$
surviving, is given by
\begin{eqnarray}
\sigma_{str}=\frac{1}{2j+1}\int d\vec{b}\,\sum_m\, \langle\phi_{0m}
\vert \vert S_2\vert^2[1-\vert S_1\vert^2] \vert\phi_{0m} \rangle.
\label{strip}
\end{eqnarray}
This cross section is seen to vanish if the interaction $V_{1T}$ of
constituent 1 with the target is non-absorptive, and hence $|
S_1|=1$.

\subsection{Time-dependent methods}
\label{timedep1}

A number of other semi-classical few-body reaction models have been
developed and applied to reactions in which the projectile is
treated as a core+valence nucleon system. One method, developed by
Bonaccorso and Brink, is to solve the time-dependent Schr\"{o}dinger
equation after assuming that the relative motion between the
projectile's core and the target can be treated classically and
approximated by a constant velocity path. This method
\cite{bon1,bon2} treats the time dependence of the reaction
explicitly and thus conserves energy, but not momentum. Breakup
amplitudes can then be calculated within time dependent perturbation
theory, referred to as the Transfer to the Continuum (TC) model
\cite{bonaccorso1}.

Another time dependent approach, described by Melezhik and Baye
\cite{melezhik}, also treating the projectile-target relative motion
semi-classically, is to solve the time dependent Schr\"{o}dinger
equation using a non-perturbative algorithm on a three-dimensional
spatial mesh that allows the treatment of Coulomb breakup in the
nonperturbative regime.

\section{Reaction cross sections}

One of the first observables measured in the study of neutron-rich
(halo) nuclei was their total interaction cross section
\cite{tan1,tan2}. This was the first indication of their extended
matter radii due to the long range tail in their neutron densities.
Theoretically, one calculates the total reaction cross section using
equation (\ref{reacsig2}). Early estimates of the size of neutron
rich isotopes of lithium and helium employed the optical limit of
the Glauber model \cite{alkhazov74} in which the nuclear one-body
densities were taken to be simple Gaussians. This allows for a
simple analytical expression to be derived \cite{karol}. This
predicted an enhanced size for these nuclei compared with that
obtained from the usual $\langle r^2 \rangle^{1/2}
\propto A^{1/3}$ scaling.

\begin{figure}
\begin{center}\includegraphics[%
	width=1.0\columnwidth]{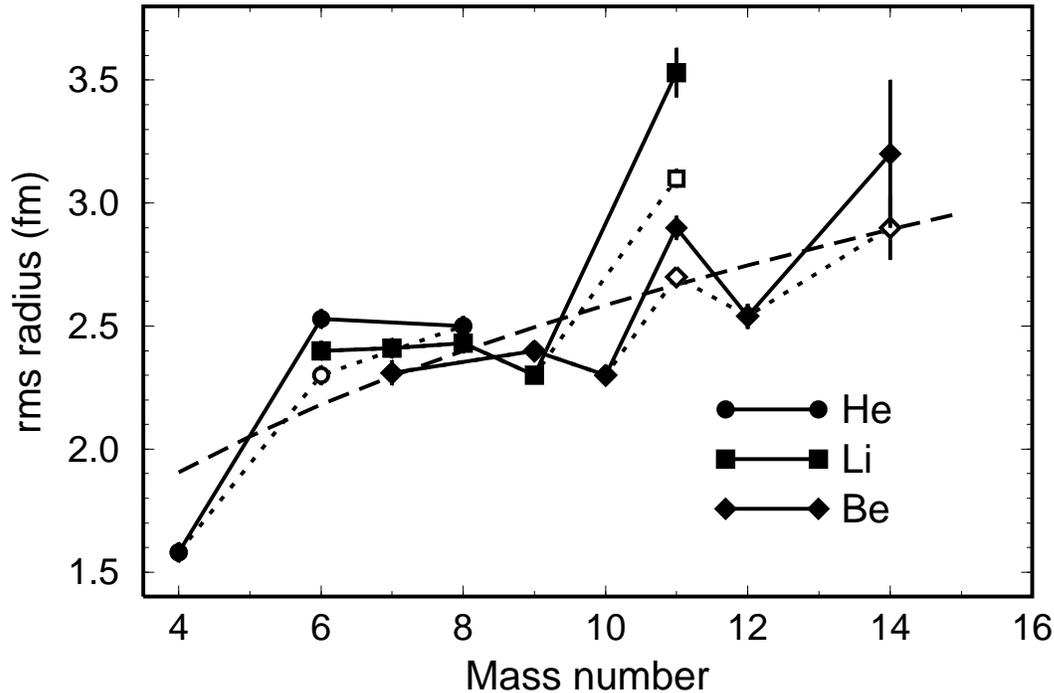}
\end{center}
\caption{\label{fig:radii}A plot of the matter radii of the
isotopes of He, Li and Be as predicted by reaction cross section
calculations \cite{alkha,alkha2} that are compared with
experimentally measured interaction cross sections. The dashed line
shows the $R_0A^{1/3}$ dependance ($R_0=1.2$fm) expected of mean
field nuclei. The open symbols, joined by the dotted lines, are the
radii predicted using one-body density description of the nuclei
within an optical limit approximation. The full symbols are
predictions using cluster wavefunctions within a few-body Glauber
approach. The open symbols are from the optical limit Glauber model
involving one-body densities derived from the same cluster
wavefunctions. In both cases, the reaction cross section is fixed to
the experimental value.}
\end{figure}

By retaining the few-body degrees of freedom in the projectile wave
function, its important structure information remains entangled. As
a consequence, studies that evaluated the reaction cross section in
equation (\ref{reacsig2}) correctly \cite{alkha,alkha2}, rather than
take the optical model limit, predicted an even larger matter
radius, as shown in figure (\ref{fig:radii}). This may at first
sight seem contrary to what we might expect, since such a model
allows for new breakup channels to become available predicting a
larger reaction cross section and hence a smaller radius to bring
the cross section back down to the experimental value again.
However, a simple yet powerful theoretical proof, due to Johnson and
Goebel \cite{inequality}, shows that for a given halo wave function,
the optical limit model always {\it overestimates} the total
reaction cross section for strongly absorbed particles, thus
requiring a smaller halo size than suggested by the full few-body
calculation for a given cross section.

\section{Elastic and inelastic scattering}

Much has been learned about the structure of light exotic nuclei from
elastic scattering, whether the nucleus of interest is scattered from a
stable nucleus or single proton. The latter case is, of course, just
proton elastic scattering in inverse kinematics. Over the past decade,
a number of measurements of the angular distribution for the scattering
of exotic weakly-bound light nuclei from a stable target (often
$^{12}$C) were unable to distinguish between elastic and inelastic
scattering due to the poor energy resolution in the detectors. Such
'quasielastic' cross sections were thus unable to resolve low-lying
excited states of the target (e.g. the $2^+$ and $3^-$ states of
$^{12}$C) from the elastic channel and the data were an incoherent sum
of elastic and inelastic pieces.

Angular distributions have been measured for $^6$He \cite{alkhahe6},
$^8$He \cite{tosthe8}, $^8$B \cite{pecina}, $^{11}$Li
\cite{kolata,lewitowicz} and $^{14}$Be \cite{zahar}. All these
nuclei are very weakly-bound and have a well-defined few-body
cluster structure. Indeed, most have only one bound state and any
excitation during the scattering process will therefore couple to
the breakup channels. Similarities were quickly drawn between these
and well-studied examples such as the deuteron (p+n), $^6$Li
($\alpha$+d) and $^7$Li ($\alpha$+t), the scattering of which is
strongly influenced by their dynamic polarization. For such
projectiles, simple folding models based on single particle
densities fail to generate the optical potentials needed to describe
the elastic scattering angular distributions.

\begin{figure}
\begin{center}\includegraphics[%
	width=0.8\columnwidth]{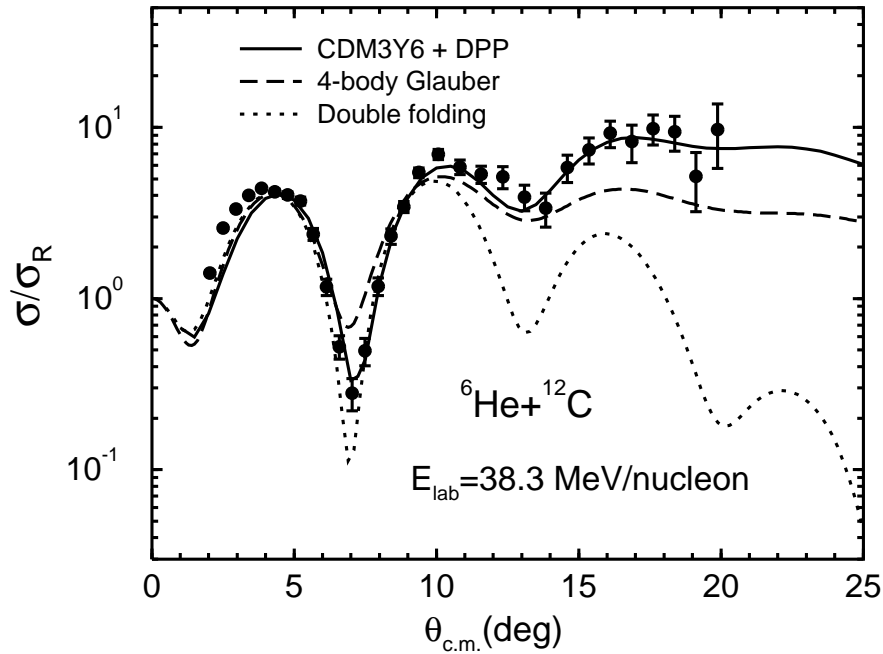}
\end{center}
\caption{\label{fig:lapoux}Elastic scattering of $^6$He$+^{12}$C at
38.3 MeV/nucleon. The data are from \cite{lapoux}. The solid curve
is an optical model fit to the data comprising of a folded
potential, using the energy- and density-dependent effective $NN$
interaction CDM3Y6 of \cite{khoa} folded over simple one-body
densities for projectile and target, plus a complex dynamic
polarization potential to account for $^6$He breakup. [See
\cite{lapoux} for further details.] The dashed curve is obtained
from a completely parameter-free four-body Glauber calculation with
a Faddeev wavefunction for $^6$He \cite{alkhahe6}. The dotted curve
was obtained by folding a $^6$He density \cite{sagawa} and a
2-parameter Fermi density for $^{12}$C with the density-dependent
DDM3Y interaction \cite{satchlerlove}.}
\end{figure}

For halo nuclei where the binding energies are typically of the order
of 1 MeV or less, the breakup effect is even more important. The
elastic scattering data for $^6$He+$^{12}$C have been analyzed within
an optical model approach, with the real part of the optical potential
calculated in the double-folding model using a realistic
density-dependent NN interaction and the imaginary part taken as a
standard Woods-Saxon form. The projectile density used in the folding
is calculated from realistic few-body wavefunctions. Such a 'bare'
folding potential, however, describes the no-breakup scattering in
which the projectile remains in its ground state throughout. An
additional phenomenological dynamic polarization potential (DPP) must
therefore be added to it to account for coupling to the breakup
channels \cite{lapoux} (see figure (\ref{fig:lapoux})).

A more microscopic approach to elastic scattering is to use a
few-body scattering model in which the few-body correlations of the
projectile are retained. In such an approach it is the few-body
wavefunction of the projectile that is used directly rather than its
one-body density. Three-body models used previously to study the
scattering of deuterons and $^{6,7}$Li have been based on CDCC,
adiabatic and Glauber approaches.  The required inputs to all these
models, in addition to the projectile few-body wavefunction, are the
projectile constituent-target optical potentials. In this way,
breakup effects are included in a natural way. In particular, one of
the advantages of the Glauber approach is that breakup is included
in a natural way to all energies and angular momenta, and to all
orders in breakup, through a closure relation. In fact, it has been
found that higher order breakup terms, such as those responsible for
continuum-continuum coupling, are indeed very important
\cite{alkhabu}.

The DPP representation of this breakup effect on the elastic channel
has been analyzed for various halo nuclei such as $^6$He and $^{11}$Li
\cite{lapoux,alkhabu,dpp,rusek,lapouxplb}. It is found to be strongly
absorptive and with a significant repulsive real part in the far
surface region, which acts to reduce the far-side scattering
amplitude.

Most few-body models have been developed to describe the scattering
of two-body projectiles (three-body scattering models). However,
many have been extended to four-body models in order to describe the
scattering of projectiles which are themselves modeled as three-body
systems, such as $^{11}$Li. First, a four-body Glauber model, based
on eikonal and adiabatic methods, was presented \cite{yabana} and
subsequently extended to include recoil and few-body correlation
effects \cite{att}. Soon after, a four-body adiabatic model - which
was fully quantum mechanical in that it made no semi-classical or
eikonal assumptions - was developed \cite{christ} based on the
three-body model of Johnson and Soper \cite{adia}. At the time of
writing this review, work is underway on a four-body CDCC
calculation. Ultimately however, the simplicity of the Glauber
approach makes it the most practical tool for describing the
scattering of projectiles composed of more than three constituents.
Using random sampling (Monte Carlo) integration, it has been
extended \cite{he8} to describe the scattering of $^8$He from
$^{12}$C in which the projectile is described by a five body
($\alpha$+4n) harmonic oscillator based Cluster Orbital Shell Model
Approximation (COSMA) wavefunction \cite{cosma}.

\begin{figure}
\begin{center}\includegraphics[%
	width=0.5\columnwidth]{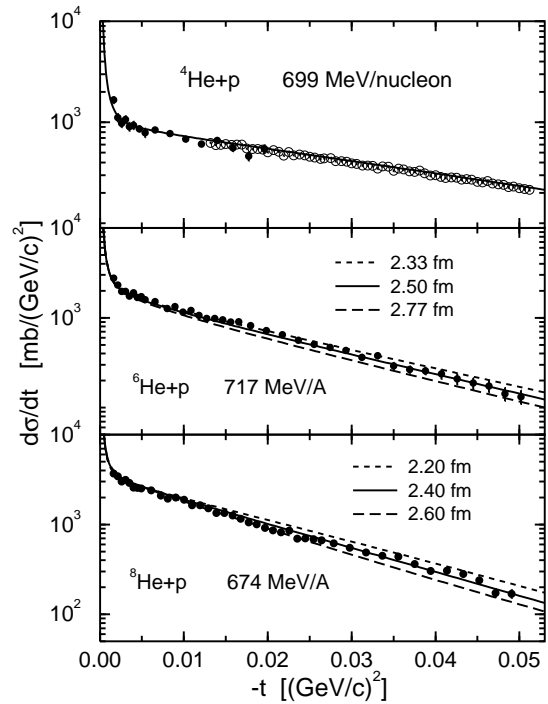}
\end{center}
\caption{\label{fig:he468}Elastic scattering of $^{4,6,8}$He from
protons at intermediate energies. At small momentum transfers such
as these the cross sections depend on the matter radii chosen for
the projectiles. The curves for $^6$He and $^8$He scattering were
obtained within a few-body Glauber model \cite{he68}. The data are
from The IKAR collaboration \cite{alkhazov1,alkhazov2}.}
\end{figure}

An analysis of high energy (700 MeV) elastic scattering of protons
from helium isotopes, $^6$He and $^8$He, in inverse kinematics has
been carried out \cite{he68,alkhazov2} to estimate their matter
radii (figure (\ref{fig:he468})). Using the Glauber model to
determine the forward angle differential cross section it was found
that while few-body correlation effects were not important at the
small momentum transfers of the experimental data \cite{alkhazov1},
nevertheless the asymptotic behaviour of the few-body wavefunctions
describing the ground states of these nuclei lead to long-range
tails in the one-body density distributions, particularly for
$^6$He. Simple analytical expressions for the densities do not give
rise to such long tails and hence under-predict the matter radius of
$^6$He by about 10\%.

\begin{figure}
\begin{center}\includegraphics[%
	width=0.8\columnwidth]{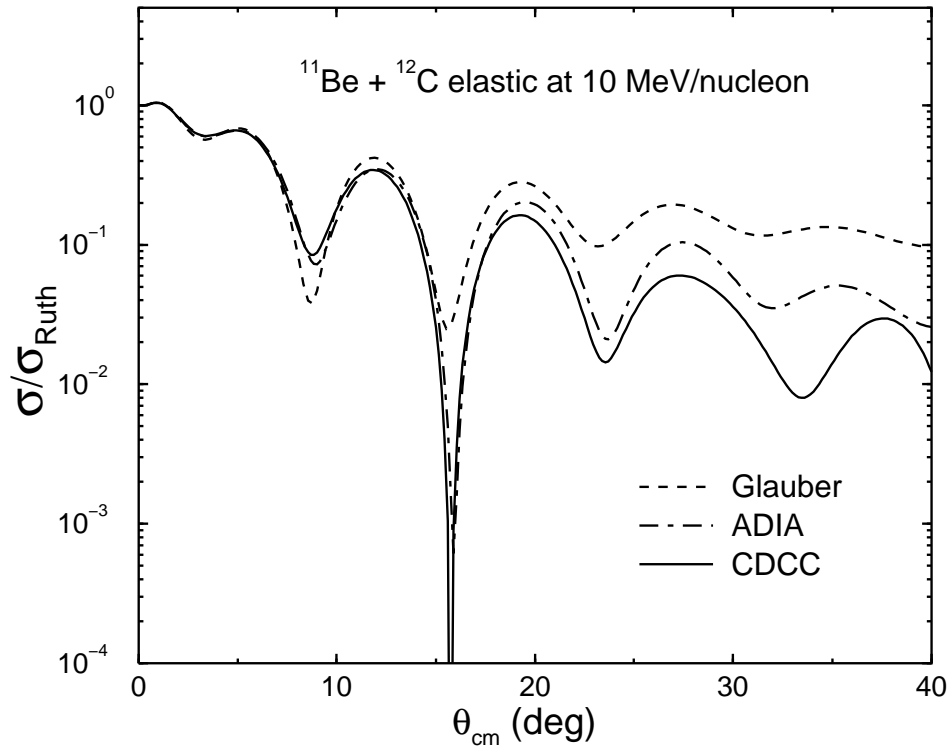}
\end{center}
\caption{\label{fig:be11elas}Calculated differential cross sections
for the elastic scattering of $^{11}$Be from $^{12}$C at 10
MeV/nucleon using various three-body models: Glauber (dashed curve),
adiabatic (dot-dashed curve) and CDCC (solid curve)
\cite{brooke,summers}. The importance of both non-adiabatic and
non-eikonal corrections at larger scattering angles can be seen
clearly at these relatively low scattering energies. No data exist.}
\end{figure}

An important drawback of Glauber methods is that they are only
accurate for forward angle, high energy, scattering. At lower
energies, corrections to the two basic assumptions (the eikonal and
the adiabatic) of the few-body Glauber model are necessary
\cite{brooke,summers}. Figure (\ref{fig:be11elas}) shows the ratio
to the rutherford angular distribution for $^{11}$Be elastic
scattering from carbon at 10  MeV/nucleon, using various models. The
solid curve is the CDCC cross section and represents here an 'exact'
calculation; the dot-dashed is also from a fully quantum mechanical
calculation but having made an adiabatic approximation; while the
dashed curve is from a three-body Glauber model calculation which
makes, in addition to adiabatic assumption, an additional
semi-classical (eikonal) approximation. Clearly, while this energy
is rather low for either the eikonal or adiabatic assumptions to
hold, both can be corrected for \cite{brooke,summers} with the
inclusion of non-eikonal and non-adiabatic terms in the elastic
amplitude.

A useful approximation to the adiabatic model, referred to as the
`recoil limit' model \cite{scatt,rcjprl}, is obtained when the
potential between the projectile's core and the target dominates
over that between the valence nucleon(s) and the target. In this
limit, when all but one of the potentials in the sum in equation
(\ref{potsum}) can be neglected, it can be shown that there is an
exact closed form to the few-body adiabatic wavefunction of equation
(\ref{adeqn}). This leads to a factorised expression for the
scattering amplitude, into a point amplitude that is, to a good
approximation, that of the core-target system at the same energy per
nucleon and momentum transfer as the full projectile, and a form
factor containing all the information on the structure and
excitations of the projectile. Such a simple formulation has proved
to be extremely useful not only in providing physical insight but as
a check of more complete coupled channels methods such as the
adiabatic and CDCC models.

The advantage of the few-body scattering models described here
is that they require two types of inputs which should,
in principle, be well-known: (i) the few-body wavefunctions that
describe the structure of many of the loosely-bound, particularly
light halo nuclei so well and (ii) information about the individual
scattering of the projectile clusters from the target, either as
optical potentials or as scattering amplitudes. It is
now well-established that the few-body dynamics should be
incorporated into the scattering and reaction mechanisms {\it ab
initio}. This necessary entangled approach goes beyond simply
feeding in knowledge of the total matter density distribution of the
projectile.

Another approach that takes into account the few-body nature of
scattering of halo nuclei is to treat it within a few-body multiple
scattering approach. Such a model, known as the Multiple Scattering
expansion of the total Transition amplitude (or MST), has been
developed by Crespo \cite{crespo1}. While traditional multiple
scattering expansions of the optical potential, such as the KMT
approach \cite{kmt}, rely on a mean field description that treats
all nucleons (in the projectile and target) on an equal footing,
they are inappropriate for exotic loosely-bound nuclei that are far
from stability. The MST approach, however, surpasses such mean field
expansions since it takes into account structure features of the
projectiles beyond the total matter density distribution alone. It
has been applied to proton elastic scattering from $^{11}$Li, in
inverse kinematics, and shows a much better agreement with data than
the KMT optical model approach due to its inclusion of both recoil
effects (of the $^9$Li core) and breakup effects of the halo
neutrons \cite{crespo2}.

Proton inelastic scattering from halo nuclei has been used as a tool
to search for evidence of the low-lying excited states in the
continuum. So far, coupled channels methods such as CDCC, which
explicitly expand on continuum states, can only be used, as
mentioned earlier, to study projectiles comprising two clusters. But
many exotic nuclei present a core + 2n three-body structure and thus
require for the time being a different approach. The reaction
$^{11}$Li$(p,p')^{11}$Li$^*$ has been analysed within both the
shake-off approximation \cite{shakeoff}, in which only the
proton-core contribution to the single scattering term is
considered, and the multiple scattering MST approach \cite{crespo3},
in which the scattering of the proton from the halo neutrons is also
taken into account.

A microscopic model, referred to as the $g$ folding potential has
been developed by Dortmans and Amos \cite{dort} who, together with
Karataglidis \cite{Karatpp}, applied it to both elastic and
inelastic proton scattering on a range of both stable and unstable
nuclei. The optical potential for the model is obtained by folding a
complex energy- and density-dependent effective $NN$ interaction
over the one-body density matrix elements and single particle bound
states of the target generated by shell model calculations. They
show that the analysis of inelastic data is a sensitive test of
nuclear structure. For instance, it has been shown \cite{lagoy} that
good agreement with data can be achieved for the inelastic
scattering to the unbound $2^+$ state at 1.87 MeV of $^6$He from
protons at 41 MeV/A, provided a large enough shell basis is used to
calculate the wavefunctions for the initial and final states of
$^6$He.

\section{Breakup reactions}
\label{bu}

Given their very low binding energy, breakup cross sections of
exotic nuclei are generally quite large and relatively easy to
measure. Consequently,  numerous breakup measurements have been
performed, even when the radioactive beam intensity was rather low.
In parallel, the theoretical community has been attempting to model
these reactions accurately. In the following sections we discuss the
results for several approaches available in the literature.

\subsection{Time-dependent calculations (Semi-classical and
Glauber approaches)}
\label{timedep}

The semiclassical theory for Coulomb excitation was developed in the
early days of nuclear physics \cite{alder}. The semiclassical
approach is valid for large impact parameters and relies on the fact
that the relative motion between the projectile and target can be
treated classically whilst the excitation of the projectile is
treated quantum mechanically. Then, the total breakup cross section
is a product of the Rutherford cross section by the square of the
excitation amplitude. The excitation amplitude is typically
calculated perturbatively, and often only E1, M1 and E2
contributions are sufficient. A further extension of this work for
relativistic energies, where the projectile follows straight line
trajectories, can be found in \cite{winther}.

The pioneering work by Baur and Bertulani \cite{baur1}, proposing
Coulomb dissociation as a source of information for radioactive
capture rates relevant in astrophysics, justifies the large efforts
that concentrated on this topic over the past decade and in
particular on the breakup of $^8$B. (For more detail, a topical
review on this subject can be found in \cite{baur2}). Calculations
in \cite{baur1} show the kinematical regimes where the breakup of
$^7$Be$\rightarrow \alpha + ^3$He  and $^{16}$O$\rightarrow \alpha +
^{12}$C on $^{208}$Pb would become useful for astrophysics (both
reactions rates have meanwhile been measured using the Coulomb
dissociation method). More relevant for our topic is the application
to $^8$B. In \cite{bertulani1}, calculations for the E1 and E1+E2+M1
are performed and compared with the RIKEN data \cite{moto}.
Controversy on the importance of the E2 contribution for this
reaction was raised due to the re-analysis of the data in
\cite{langanke}. Improvements to the reactions models, which
will be covered in the following subsections, have shown that the
quadrupole contribution is not easy to disentangle.

Higher order corrections to include the two photon exchange was
developed in \cite{typel1} and corrections up to third order for the
Coulomb interaction was deduced within the small adiabaticity
approximation \cite{typel2}. The second order correction is always
positive but the third order interferes to produce the so called
dynamical quenching of the E2 strength \cite{esbensen2}. These
corrections are more important at lower beam energies and for small
relative energies of the fragments resulting from the breakup of the
projectile. In addition, nuclear diffraction effects need also be
considered. The semiclassical approach neglects the strong
interaction between the projectile and the target. Diffraction
corrections on the Rutherford orbit were calculated by comparing the
semiclassical expressions with the Glauber approach \cite{typel2}.
These effects can become very large even for small angles.

Although the first order semiclassical method is appealing due to
its simplicity, there are many aspects of the problem that are left
out. One of the debated issues concerned the post-acceleration of
the light fragment in the Coulomb field. In order to describe this
process properly, one should formulate the problem
non-perturbatively.

\begin{figure}
\begin{center}\includegraphics[%
	width=1.0\columnwidth]{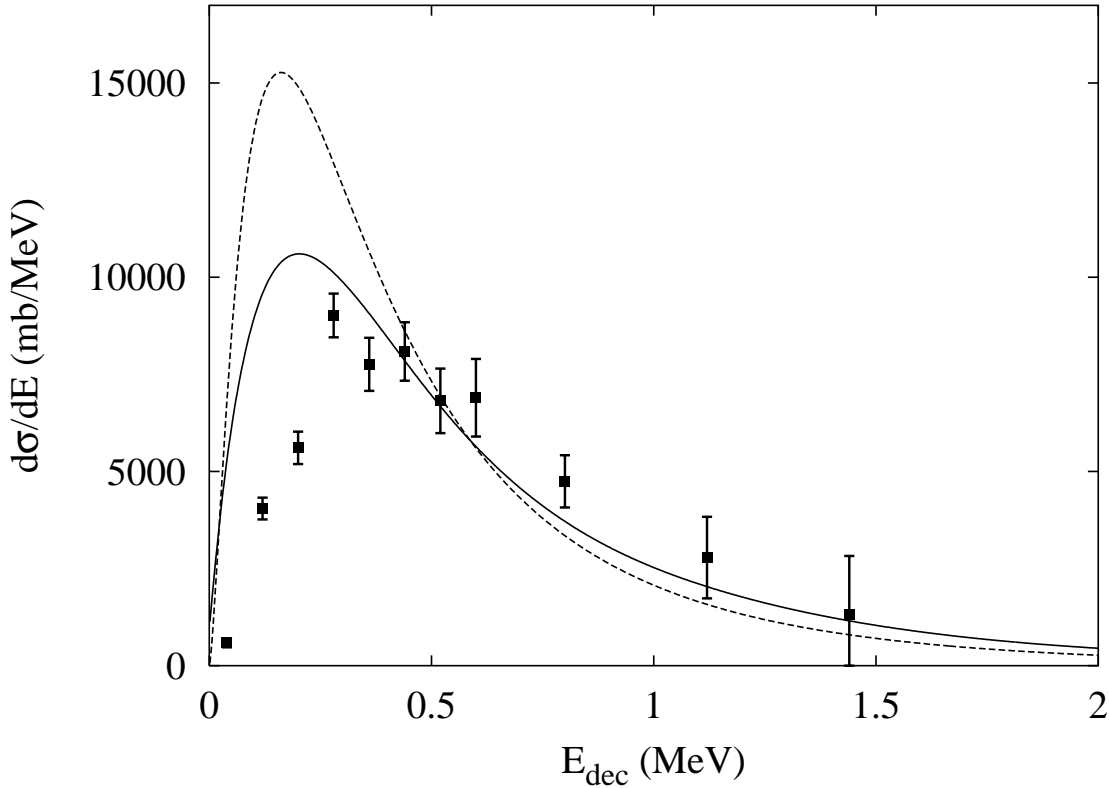}
\end{center}
\caption{\label{fig:esbensen}The breakup of $^{11}$Li on a lead
target at 28 MeV/A: first order perturbation theory (dotted line)
versus the dynamical calculation (solid line). The calculations are
from \cite{esbensen1} and the data is from D. Sackett {\it et al.},
Phys. Rec. C 48 (1993) 118.}
\end{figure}

Instead of treating the time dependent Hamiltonian perturbatively,
an exact treatment can be performed. This approach, known as the
dynamical method and introduced in section \ref{timedep1}, has two
great advantages: i) it contains the coupling to breakup channels to
all orders and ii) the multipole expansion of the Coulomb
interaction is not necessary. In most applications to data, the
projectile is still confined to the Rutherford trajectory. Such
applications include the breakup of $^{11}$Li (using a dineutron
model) and $^{11}$Be \cite{esbensen1,romanelli}. The time evolution
of the projectile wave function was calculated by solving the
three-dimensional time dependent Schr\"odinger equation. In both
examples the dynamical calculation reduces the cross section when
compared with first order perturbation theory, although this effect
is more noticeable at lower energies. An example of the breakup of
$^{11}$Li on $^{208}$Pb at 28 MeV/A is shown in figure
(\ref{fig:esbensen}), comparing the first order perturbation theory
with the time dependent dynamical calculation. The re-acceleration
is produced automatically in the calculation, shifting the momentum
distributions. The comparison with the breakup data for $^{11}$Be is
good. For $^{11}$Li, the calculation produces a different energy
distribution, a limitation not of the reaction model but of the
dineutron structure model. The application of this method to the
breakup of $^8$B at intermediate energies
\cite{bertulani1,esbensen2} shows the importance of including all
the dynamics, as in this case there is a strong E1/E2 interference
that reduces substantially the breakup probability.

More recent time-dependent calculations including both the nuclear
and the Coulomb interaction between projectile and target have been
performed for the low energy breakup of $^{8}$B
\cite{vitturi,esbensen3}. In the first of these, the calculation is
partly truncated, in the sense that no continuum-continuum couplings
are included. The comparison between these two calculations shows
that this is not an adequate approximation for this system. Similar
calculations were performed for the breakup of $^{17}$F, another
exotic nucleus on the proton dripline \cite{esbensen4}. Conclusions
are qualitatively similar to those for $^8$B.

Other implementations of the non-perturbative time dependent
approach were performed by \cite{melezhik} and \cite{fallot}. The
Lagrange-mesh method of \cite{melezhik} used the breakup of
$^{11}$Be on $^{208}$Pb at 72 MeV/A as a test case. A careful study
of the convergence of the method demonstrates that the equation
needs to be solved for a radial grid within a large range
($R_{max}=1200$ fm) keeping the radial step very small. Deflection
from the straight line trajectories and a spin-orbit coupling to the
Coulomb field are included but proven to be weak for the particular
case studied. The calculations in \cite{fallot} have been successful
in analysing the recent $^{11}$Be data from GANIL.

Another possibility for calculating neutron breakup observables for
nuclei on the neutron dripline is the semiclassical TC (Transfer to
the Continuum) model \cite{bonaccorso1} described in subsection
\ref{timedep1}. In this model, the initial and final wavefunctions
are approximated to their asymptotic forms, and the WKB
approximation is made to the distorted waves describing the
projectile-target relative motion, providing a simple analytic
expression for the Coulomb breakup amplitude. It is generally
applicable to reactions at large impact parameters and intermediate
energies. It has been demonstrated that this model reduces to the
PWBA (Plane Wave Born Approximation) when the binding energy of the
projectile tends to zero, and to the Serber formula in the high
energy limit.

Often, an independent treatment of the nuclear and Coulomb parts is
preferred. For example, in \cite{bonaccorso2} a semiclassical model is
developed to study interference effects in the breakup of one-neutron
halos. Results for $^{11}$Be are calculated when it reacts with three
separate targets. The Coulomb breakup contribution is calculated within
a first order semiclassical approach, whereas the nuclear
neutron-target interaction is treated to all orders.  These are finally
added incoherently. More recently, the same authors have presented a
model in which they calculate both Coulomb and nuclear breakup to all
orders consistently within an eikonal framework \cite{bonaccorso3}.

Unfortunately, even though it is now generally known that dynamical
effects are very important, often the first order semiclassical
theory is still used (e.g. the dissociation of $^{19}$C
\cite{nakamura1} or $^{8}$He \cite{iwata}).

\subsection{DWBA calculations}
\label{dwba}

A traditional quantum mechanical approach to the breakup reaction
uses distorted waves for the initial and final states of the
relative motion between the projectile and the target, as well as
the Born approximation: the one-step DWBA. The nuclear part of early
RIKEN data for $^{8}$B breakup on $^{208}$Pb was analyzed using this
approach \cite{moto}, whilst the Coulomb part was treated in
first-order semiclassical theory. This reaction was re-measured with
better accuracy and angular coverage \cite{kikuchi}. These data were
re-analyzed using DWBA for both nuclear and Coulomb
\cite{shyam2,shyam3}. The results show evidence for the strong
model-dependence of the E2 contribution.

At lower energy, the E2 component becomes stronger. A series of
experiments were carried out in Notre Dame \cite{b8nd1,b8nd2,b8nd3}
with the  aim of pinning down this ingredient. The DWBA calculation
of \cite{nunes-dwba} for this system used, instead of the
conventional collective model for the coupling, the folding with the
$^8$B wavefunction. This aspect is essential for loosely bound
projectiles. A calculation for the angular distribution, for Coulomb
only, shows that the finite range effects of $^8$B become noticeable
for angles as low as $30$ degrees, much lower than what would be
expected through impact parameter considerations. In that work a
nuclear peak is predicted around $80$ degrees, which disappeared
once all couplings were included (next subsection). The importance
of Coulomb-nuclear interference is also underlined.

\begin{figure}
\begin{center}\includegraphics[%
	width=0.5\columnwidth]{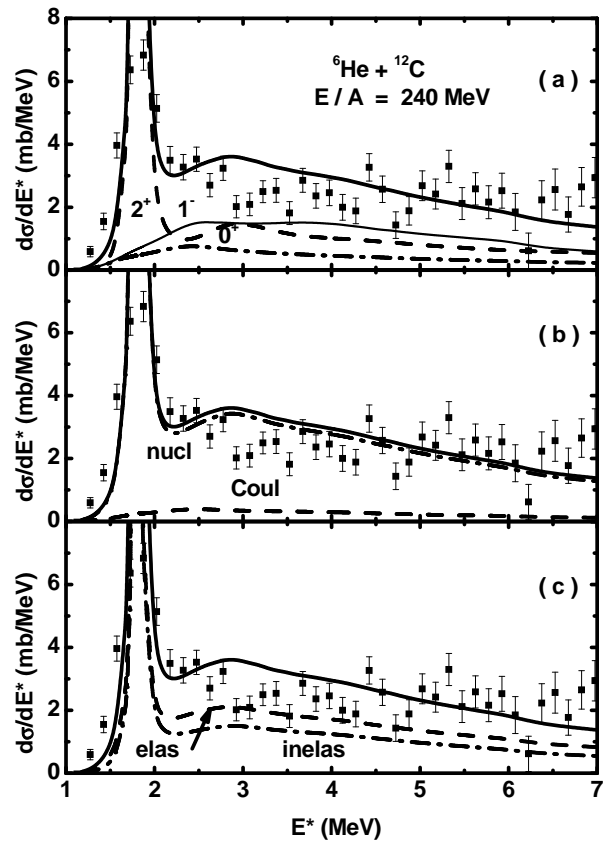}
\end{center}
\caption{\label{fig:ershov}The breakup of $^{6}$He on $^{12}$C at 240
MeV/A: a) contribution of $1^-$ (thin solid), $2^+$ (long dashed)
and $0^+$ (dot-dashed) to the differential cross section; b)
contribution of the nuclear (dot-dashed) and the Coulomb part (long
dashed); c) contribution of elastic (long dashed) and inelastic
fragmentation (inelastic). In all cases the thick solid lines
correspond to the total differential cross section and are compared
with the data from T. Aumann {\it et al.}, Phys. Rev. C 59 (1999)
1252. }
\end{figure}
All the above mentioned approaches describe the breakup of two body
projectiles. The work for calculating the three body breakup cross
section was initiated with the development of the four-body DWIA
(Distorted Wave Impulse Approximation) \cite{ershov1}. This method
offers a one step quantum mechanical calculation only valid for high
energies, when the loss of energy in the breakup is small compared
with the initial energy. The calculations were applied to $^6$He
breakup in the fields of $^{12}$C and $^{208}$Pb. Results show  that
including the full three body structure of the projectile enables a
very rich interplay between the reaction mechanism and the halo
excitations that otherwise would be missing (an illustration is
given in figure \ref{fig:ershov}). The major drawback of this model
is that the four body partial wave expansion is extremely
cumbersome. Preliminary four-body DWBA calculations for the Coulomb
breakup of $^6$He have also been presented in  \cite{chatterjee}.

\subsection{CDCC calculations}
\label{cdcc2}

The CDCC method \cite{cdcc1,cdcc2} briefly introduced in section
\ref{cdcc}, offers one of the most complete approximation to the
three body problem involving a 2-body projectile impinging on a
target. It has been shown that the exact Faddeev equations reduce
to the CDCC equations as long as the model space is sufficiently
large \cite{cdcc-fad}. And even though convergence issues need to be
carefully checked, solving the CDCC equations is much easier than
finding the Faddeev solutions to the problem.

The CDCC method reduces to the DWBA when only one-step processes
are taken into account.  One can further solve the coupled channel
equations iteratively, including 2,3,...,n steps in the reaction.
This method should obviously converge to the CDCC exact solution.

Before discussing specific applications of the CDCC method, we
emphasize that it is not always trivial to obtain a model space
which is sufficient to account for all the physical properties (e.g.
\cite{kin3b}). Convergence studies concerning the choice of the
discretization were performed in Ref.\cite{cdcc-conv} for $^{58}$Ni
elastic and breakup reactions, proving that both the average method
and the midpoint method provide the same results. However, it should
be stressed that that study considered nuclear coupling only, and
not Coulomb. The main advantage of the average method is that the
resulting bin-wavefunctions characterizing the continuum states are
square integrable and thus couplings between two continuum states
are tractable.

\begin{figure}
\begin{center}\includegraphics[%
	width=0.8\columnwidth]{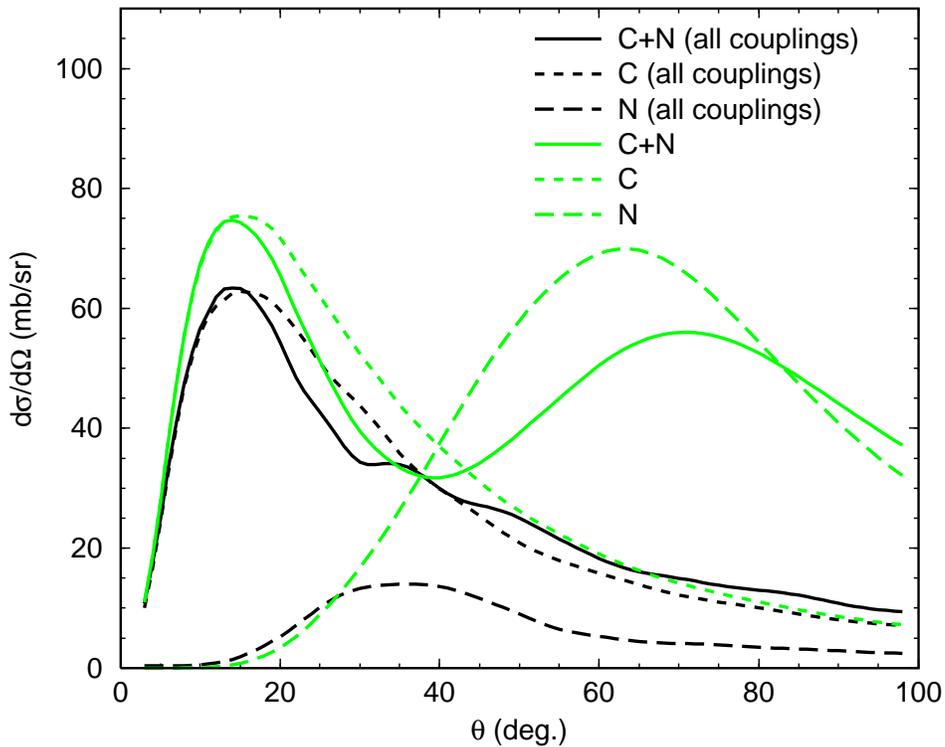}
\end{center}
\caption{\label{fig:nunes}The breakup of $^{8}$B on $^{58}$Ni at 26
MeV: nuclear only (dashed line), Coulomb only (dashed) and
the result of including both nuclear and Coulomb (solid).
The grey lines are the truncated calculations
without couplings between continuum states. }
\end{figure}
The first application of the CDCC method to the breakup of exotic
nuclei was performed for the Notre Dame experiments
\cite{b8nd1,b8nd2,b8nd3}: 25.8 MeV $^8$B, breaking into $^7$Be+p
under the field of $^{58}$Ni \cite{kin3b,nunes-cdcc}. In
\cite{nunes-cdcc} differential cross sections for multi-step
processes are calculated for both nuclear and Coulomb separately. It
is shown that even six-step processes have a significant
contribution. Here too, Coulomb and nuclear effects need to be
included coherently, as interference plays an important role. In
figure (\ref{fig:nunes}) the full CDCC calculation is compared with
the truncated calculation where no continuum-continuum couplings are
included.  The huge reduction of the nuclear peak can be only
attributed to the couplings within the continuum. The three-body
observables were adequately derived in \cite{kin3b}. This piece was
required due to the fact that the data had incomplete kinematics
\cite{b8nd1,b8nd2,b8nd3} (the outgoing proton was not detected).
The calculated $^7$Be angular and energy distributions could then
be compared directly with the data. The agreement is excellent for
all but the largest detection angle, where transfer effects become
relevant. Yet even this fact can be easily accounted for within the
model \cite{b8nd3}.

\begin{figure}
\begin{center}\includegraphics[%
	width=0.8\columnwidth]{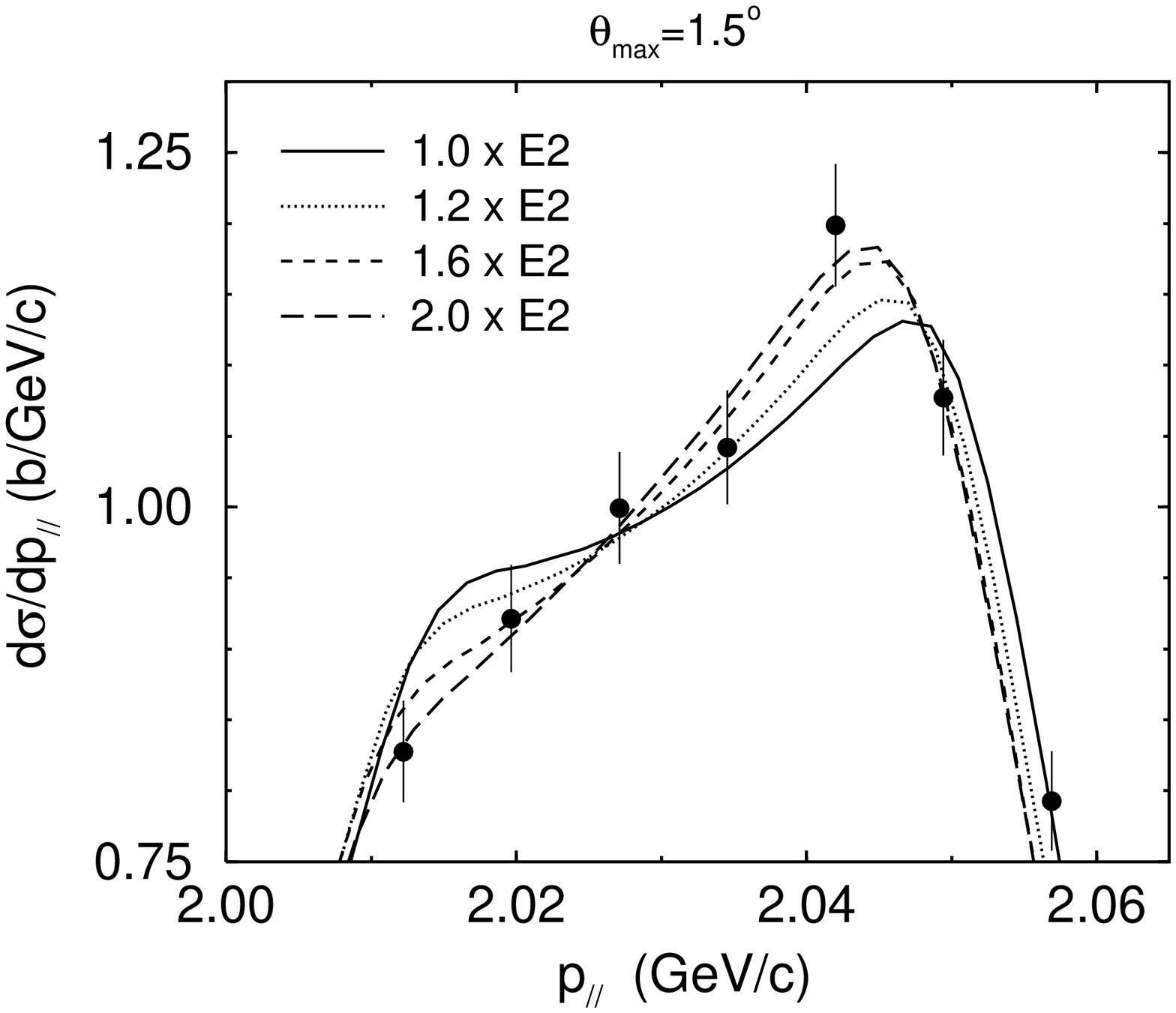}
\end{center}
\caption{\label{fig:mortimer}The breakup of $^{8}$B on lead at 44
MeV/A: the quadrupole components of the reaction process has
been multiplied by the given factor. The data is from \cite{davids3}.}
\end{figure}
More recently there has been an application of the CDCC method to
the breakup of $^8$B at higher energies \cite{mortimer}. In the
previous years, a series of $^8$B Coulomb dissociation experiments
were performed at MSU \cite{davids1,davids2,davids3}. Breakup data
on both Pb and Ag targets, at 44 MeV/u and 81 MeV/u, were compared
with CDCC calculations in \cite{mortimer}. Therein it is shown that
the asymmetry of the momentum distributions is reduced through the
couplings to the continuum, in agreement with the results at lower
energy \cite{kin3b}. Also, in order to obtain agreement with the
data, it was found that the quadrupole component needed to be scaled
by 1.6, as shown in figure (\ref{fig:mortimer}).

The motivation for both the Notre Dame and the MSU series of
measurements on $^8$B is astrophysical. As mentioned earlier,
Coulomb dissociation may offer a powerful tool for extracting
S-factors, as long as the E2 component can be well determined,
nuclear effects are negligible and no higher order effects are
present. Both Notre Dame and MSU experiments aimed at pinning down
the E2 components, that are part of the dissociation cross section
but do not contribute to the direct capture transition at low
energy. The data from Notre Dame \cite{b8nd3} is consistent with the
modified $^8$B model  from Esbensen and Bertsch \cite{esbensen2}
(where all breakup states are calculated with the ground state
single particle interaction). However, as just mentioned, the MSU
data \cite{davids3} requires a 1.6 increase of the E2 component. It
is not clear whether the inconsistency comes from the reaction model
or missing structure information.

CDCC calculations for the breakup of $^6$Li and $^7$Li on Pb have
been used to analyze the recent data from Florida State
\cite{kelly}. Experiment shows that the $\alpha$-breakup for $^6$Li
is systematically larger than for $^7$Li. The CDCC results predict
the correct trend in the measured energy range (29 MeV to 52 MeV)
although the absolute value is considerably lower than expected. The
most likely reason pointed out by the authors for this mismatch
\cite{kelly} is  the absence of transfer channels in the
calculation. This was also found to be important for $^6$He
\cite{he6nd} and we will come back to this point in section
(\ref{transfer}).

\subsection{Other approaches to Breakup}
\label{otherbu}

Although in the previous subsections we have covered the main
theoretical approaches to breakup, some alternative methods,
that were developed with specific applications in mind,
should also be mentioned.

The adiabatic method described in Subsection \ref{adiamodel} is
particularly interesting for neutron rich nuclei
\cite{supa,banerjee}. This method involves a fully
quantum-mechanical non-perturbative description of the pure Coulomb
breakup process, where essentially two approximations are made: i)
the valence particle(s) of the projectile does(do) not interact with
the target and ii) the relative motion of the fragments in the
projectile is treated adiabatically. This method is thus only
applicable to projectiles with a neutral valence particle (such as
$^{11}$Be) and for reactions performed at relatively high energy. In
\cite{supa} results for the breakup of the deuteron are compared
with various sets of data, and a good level of agreement is
obtained. The application to three body projectiles is made in
\cite{banerjee}. The comparison  of the theory with the experimental
data for the breakup of $^6$He shows that for this reaction
\cite{balamuth}, nuclear effects are considerable and finite range
effects on the Coulomb interaction need to be taken into account.

Even though the equivalence between the full Faddeev formalism and
the CDCC truncation has been proven \cite{cdcc-fad}, in practice the
CDCC calculations are truncated in configuration space and matched
to two body asymptotics. Work developed by Alt and Mukhamedzhanov
\cite{alt} estimated the correction to these approximations,
handling the Faddeev two- to three-body scattering problem. The
improvement is that the asymptotics for the  three charged particles
final state is correctly included. The application to the breakup of
$^8$B shows that these asymptotic effects are more important for
larger angles/larger relative energies. They are not relevant for
the high energy GSI experiment \cite{iwasa} but should not be
neglected in the RIKEN data \cite{moto,kikuchi}.

The participant spectator model (PSM) has been proposed to calculate
high energy reaction observables for three body projectiles
impinging on a target. It takes the sudden approximation (neglecting
completely the internal energy of the projectile) and assumes that
only one of the constituents of the three body nucleus interacts
with the target at a time. Initially applied to light targets, where
the process was nuclear dominated \cite{garrido1,garrido2}, it has
meanwhile been extended to treat Coulomb processes too
\cite{garrido3}. A wide variety of observables are computed. For the
$^{11}$Li data, it seems to be possible to choose reasonable radial
cutoff parameters within a black disk approximation, for each
fragment separately, that provide an overall agreement.

In the data analysis of many three body breakup experiments,
the mechanism is often interpreted as a decay through the
existing two-body resonances of the subsystems. In \cite{garrido4}
it is shown that a correspondence between the R-matrix
resonance parameters and the real resonance structure of the
two body subsystems is not always possible.

\subsection{Momentum distributions}
\label{momdis}

Measurement of the momentum distributions of the fragments (core and
valence nucleons) following the breakup of halo nuclei on stable
targets is now a well-established method for studying halo
properties. While it has been used for many decades as a tool to
access the structure of stable nuclei, it is particularly
well-suited to loosely bound systems. The basic idea is simple:
since very little momentum transfer is required in the breakup
process to dissociate the projectile fragments, they will be
detected with almost the same velocity as they had prior to breakup,
and their relative velocities will be very similar to those within
the initial bound projectile. In all reactions with weakly-bound
systems the momentum distributions are found to be very narrowly
focussed about the beam velocity. This has the simple physical
interpretation of representing the momentum distribution of the
initial projectile. Via the Uncertainty Principle, the narrow
momentum distribution corresponds to a wide spatial distribution. It
was such observed narrow distributions which first helped confirm
the large extent of halo nuclei \cite{kobayashi}. Since then, many
measurements have been made, involving detection of both the valence
nucleons and the core fragments, and the halo structure of several
light nuclei has been established.

Two types of distributions can be measured: either perpendicular
(transverse) or parallel (longitudinal) to the beam direction. It is
now known that the transverse distributions are broadened due to
nuclear and Coulomb diffraction effects (elastic scattering of the
fragments from the target) and therefore require more careful
theoretical analyses. This is why longitudinal momentum
distributions are regarded as a better probe of the projectile
structure. Early on, simple models, based on eikonal assumptions
agreed with measurements rather well. Similar widths were obtained
from nuclear breakup on light targets and Coulomb breakup on heavy
targets, supporting the view that the distributions were no more
than the square of the Fourier Transform of the projectile ground
state wavefunction. However, presently this view is considered too
simplistic. Firstly, these simple models only really sample the
momentum content of the single particle wavefunction at the nuclear
surface \cite{hansen}. Secondly, reaction mechanisms, for the case
of two neutron halo nuclei, need to be taken into account since the
neutrons may be scattered or absorbed separately. Even for single
valence nucleon systems, such as $^8$B, the reaction mechanisms lead
to a narrowing of the calculated width due to the valence proton
being in a relative $p$-state and the $m=\pm 1$ components of the
wavefunction being affected differently in the breakup process
\cite{esbensen2,obuti}.

For single valence nucleon systems, the longitudinal inelastic
breakup momentum distributions for the core - at high energies the
elastic breakup piece is small - can be expressed as
\begin{eqnarray}
\frac{{\rm d}\sigma}{{\rm d}{k_z}}=\frac{1}{2l+1}\sum_{m=-l}^{l}
&\int d\vec{s}& \ \left\vert \frac{1}{\sqrt{2\pi}}\int dz\, e^{ik_zz}
\,\phi_{lm}(\vec{s},z)\right\vert^2 \nonumber\\
&&\times\int d\vec{b}_c\ \vert S_c(b_c)\vert^2\left( 1-\vert
S_n(b_n)\vert^2\right) ,
\label{momdist}
\end{eqnarray}
where $b_n = \vert \vec{b}_c+\vec{s}\,\vert$ and $\vec{s}$ is the
projection of the core-nucleon relative coordinate onto the impact
parameter plane, $\phi_{lm}(\vec{s},z)$ is the valence nucleon
wavefunction with orbital angular momentum $lm$ and $S_c$, $S_n$ are
the core and nucleon elastic S-matrices as described in Subsection
\ref{glauber}. The integral over $\vec{b}$ in Eq. (\ref{momdist})
represents the reaction mechanism and involves the product of the
core survival probability (in its ground state) and the nucleon
absorption probability by the target. Without this factor, the
momentum distribution is just a Fourier transform of the nucleon
wavefunction.

\begin{figure}
\begin{center}\includegraphics[%
	width=1.0\columnwidth]{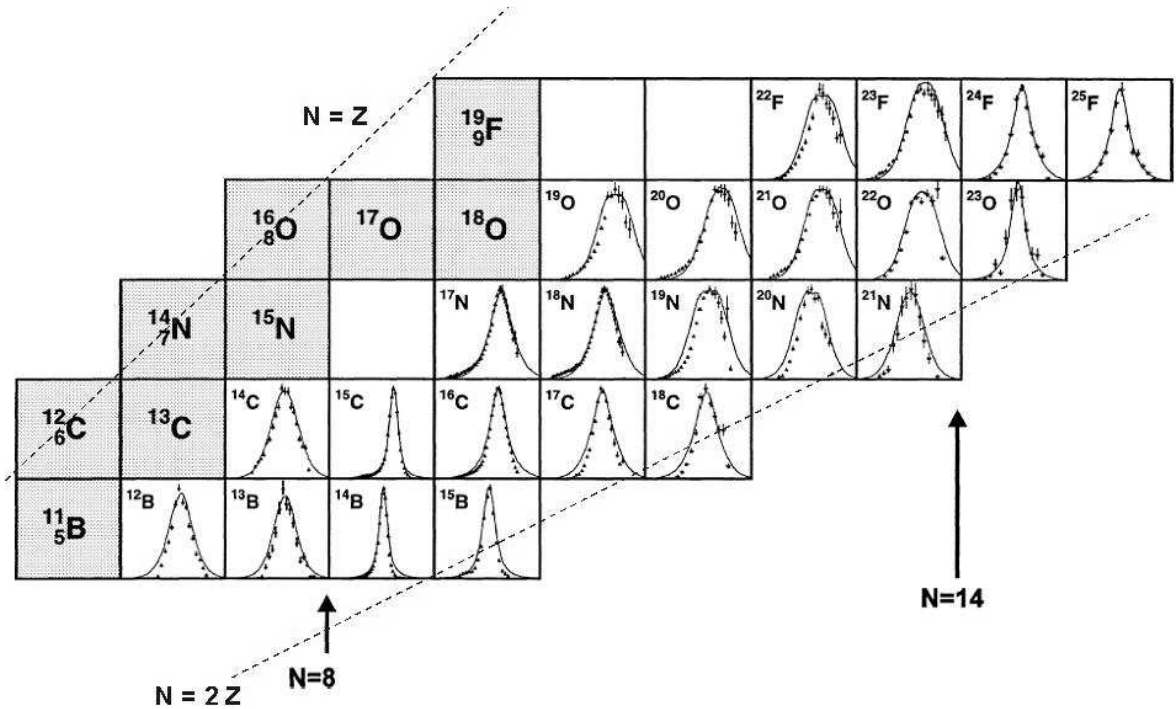}
\end{center}
\caption{\label{fig:sauvan}The longitudinal momentum distributions
for the core fragments following single neutron removal from a range
of neutron-rich nuclides on a carbon target \cite{sauvan}. The
narrow distributions correspond to larger size. }
\end{figure}

Figure (\ref{fig:sauvan}) shows a segment of the segre chart with
the momentum distributions for individual nuclides superposed. The
theoretical curves were obtained using a Glauber model with JLM
parameterisation of the optical potentials and incorporating second
order non-eikonal corrections (see ref. \cite{sauvan} for details).

Another complication is the possibility of final state interactions
between the surviving fragments. These can lead to 2- or 3-body
resonances that also act to make the widths narrower
\cite{garrido4,barranco,zinser,ershov,nunes-cdcc}.

Finally, in measurements in which the surviving fragments formed a
bound state, there was a significant probability that the reaction
would populate excited states of this system. This implied that the
measured momentum distributions contained several distributions
superimposed, each with a different width. This has led to gamma-ray
coincidence measurements to discriminate between the different
partial cross sections, as will be discussed in the next section.


\section{Knockout reactions}
\label{knockout}

The early measurements of Coulomb dissociation and one nucleon
removal cross sections of halo states have since evolved into the
more general technique of single neutron knock-out reactions, which
have become a reliable tool for obtaining basic information about
the shell structure of a number of neutron-rich nuclei. The neutron
knockout process takes place via two different mechanisms:
diffractive dissociation (elastic breakup) and stripping (neutron
absorption by the target). Theoretically, each of these two
contributions is evaluated separately, usually within a Glauber
framework. In particular, the stripping cross section can be
calculated within a model in which the projectile comprises of the
stopped neutron plus the surviving fragment. Such a `three-body'
model (fragment+neutron+target) treats the detected fragment as a
'spectator core' which, at most, interacts elastically with the
target.

The spectator core assumption in models of nuclear-induced breakup
or a knockout reaction was first proposed by Hussein and
McVoy\cite{hussein} and has more recently been applied to the study
of the break-up of halo nuclei\cite{barran,bertsch} where it is
based on a few-body eikonal approach. Tostevin\cite{tost} has
proposed a modified spectator core model for the calculation of
partial cross sections to definite final states of the surviving
core fragment.

Considering first the simple case of a two-cluster (core+n)
projectile interacting with a target, the total cross section for
stripping of the neutron and detecting the surviving core ($c$) in a
particular final state, $J_c^\pi$ (spin $J_c$ and parity $\pi$), can
be written as
\begin{equation}
\sigma_{st}(J_c^\pi) = \frac{1}{2J+1}\int d{\bf b} \sum_{M} \langle
\Phi_{JM}^c\vert\ (1-\vert S_n(b_n)\vert^2)\ \vert
S_c(b_c)\vert^2\ \vert\Phi_{JM}^c\rangle\ ,
\label{tost}
\end{equation}
where $\Phi_{JM}^c$ is the ground state wave function of the
projectile, with angular momentum $J$ and projection $M$, containing
the core fragment in state $J_c^\pi$. This is thus only part of the
projectile's total ground state wave function which may well contain
contributions from configurations involving different core states.
Note that Eq. (\ref{tost}) is essentially the same as Eq.
(\ref{strip}) only here the cross section is just that part of the
full stripping cross section in which the spectator core is in the
state $J_c^\pi$ both before and after the stripping process.

\begin{figure}
\begin{center}\includegraphics[%
	width=0.5\columnwidth]{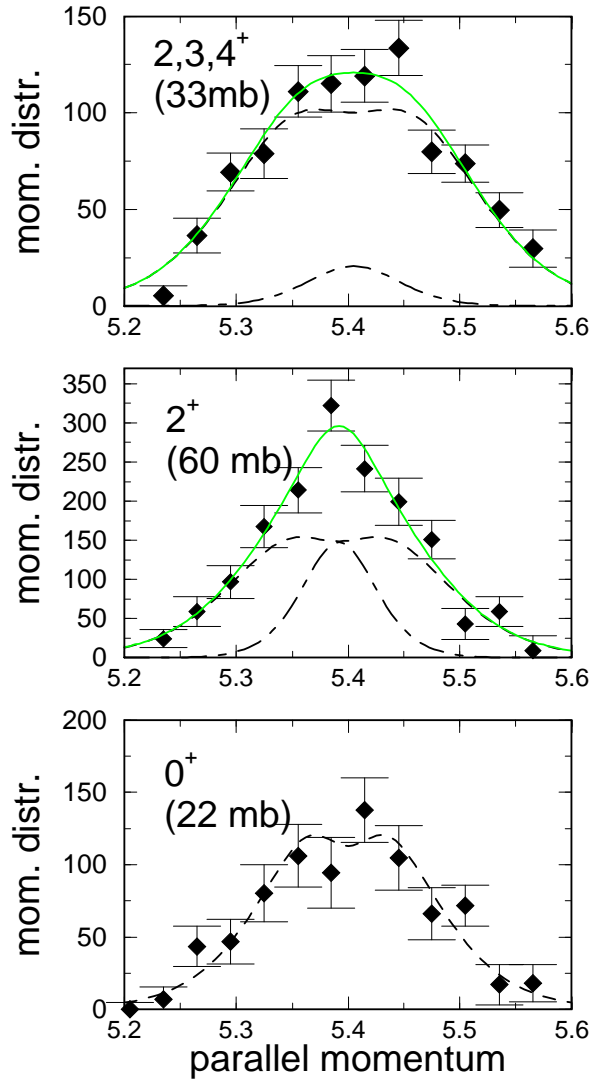}
\end{center}
\caption{\label{fig:hansen}Partial longitudinal momentum distributions
corresponding to the states in  the simplified level scheme of
$^{16}$C. The bottom panel corresponds to populating the g.s., the
middle panel to the $2^+$ state at 1.77 MeV and the top panel to the
bunch of three states all around an excitation energy of $\sim 4.1$
MeV. The data is from \cite{mad}.}
\end{figure}

As an example of the procedure we present the case for the
($^{16}$C,$^{15}$C) \cite{mad} in figure (\ref{fig:hansen}). The
momentum distribution for the outcoming $^{15}$C are measured along
with any coincidence $\gamma$-rays from the decay of excited core
states, allowing the extraction of a spectroscopic factor and
angular momentum from the overall normalisation and the shape of the
distribution.

When dealing with a three-body projectile involving a core and two
valence neutrons, the surviving fragment in equation (\ref{tost}) (i.e.
after the removal of just one of the neutrons) is
now itself a composite system of core plus neutron, which may be
loosely-bound. To check whether the spectator core assumption
remains valid in this situation, a four-body generalisation of the
Tostevin model, that allows for the dynamic coupling of different
fragment states in the stripping process, was developed \cite{evora}
and applied to a number of reactions such as
$^9$Be($^{12}$Be,$^{11}$Be$\gamma$) at 78 MeV/A. For this reaction,
partial cross sections to both the ${\frac{1}{2}}^+$ ground state
and ${\frac{1}{2}}^-$ first excited state of $^{11}$Be have been
measured and calculated \cite{navin2}. It is known that only one
third of the ground state wave function of $^{12}$Be (which is
treated as a three-body ($^{10}$Be+2n) system \cite{nunes}) comes
from a closed $p$-shell configuration, with the valence neutron pair
spending most of their time in the ($1s^2 + 0d^2$) intruder
configuration. Clearly, the $s_{1/2}$ intruder ground state in
$^{11}$Be has some effect on the configuration mixing in $^{12}$Be.
However, the spectroscopic factors deduced from the cross sections
estimated in the Tostevin spectator model will be modified if
dynamical coupling between the different $^{10}$Be+n states
($1s_{1/2}$, $0p_{1/2}$ and $0d_{5/2}$) in the projectile and the
bound states ($1s_{1/2}$ and $0p_{1/2}$) of the final $^{11}$Be are
important.

It was found \cite{evora} that allowing for couplings between
different single particle states caused a less than 10\% overall
increase to the stripping cross section. Such a correction gives an
indication of the reliability of the spectator assumption and
suggests that it is better than might be expected for such a `core'
as $^{11}$Be. Similar modifications to the partial stripping cross
sections have also been found when applied to the reaction
($^{16}$C,$^{15}$C), where the knockout cross sections to the
${\frac{1}{2}}^+$ ground state and ${\frac{5}{2}}^+$ excited state
of $^{15}$C have been measured \cite{mad}.  In this case, including
the dynamical coupling between different single particle states of
the valence neutron in $^{15}$C give rise to an overall {\em
reduction} in the stripping cross sections. The changes are
nevertheless relatively small (of order 5\%).

It should be emphasised that there are other effects which, if
included, could also affect the calculated cross section, such as core
recoil, the use of a more realistic three-body wave function for the
projectile and, maybe most importantly, including collective core
excitation effects in both the initial and final wave functions.

\begin{figure}
\begin{center}\includegraphics[%
	width=0.7\columnwidth]{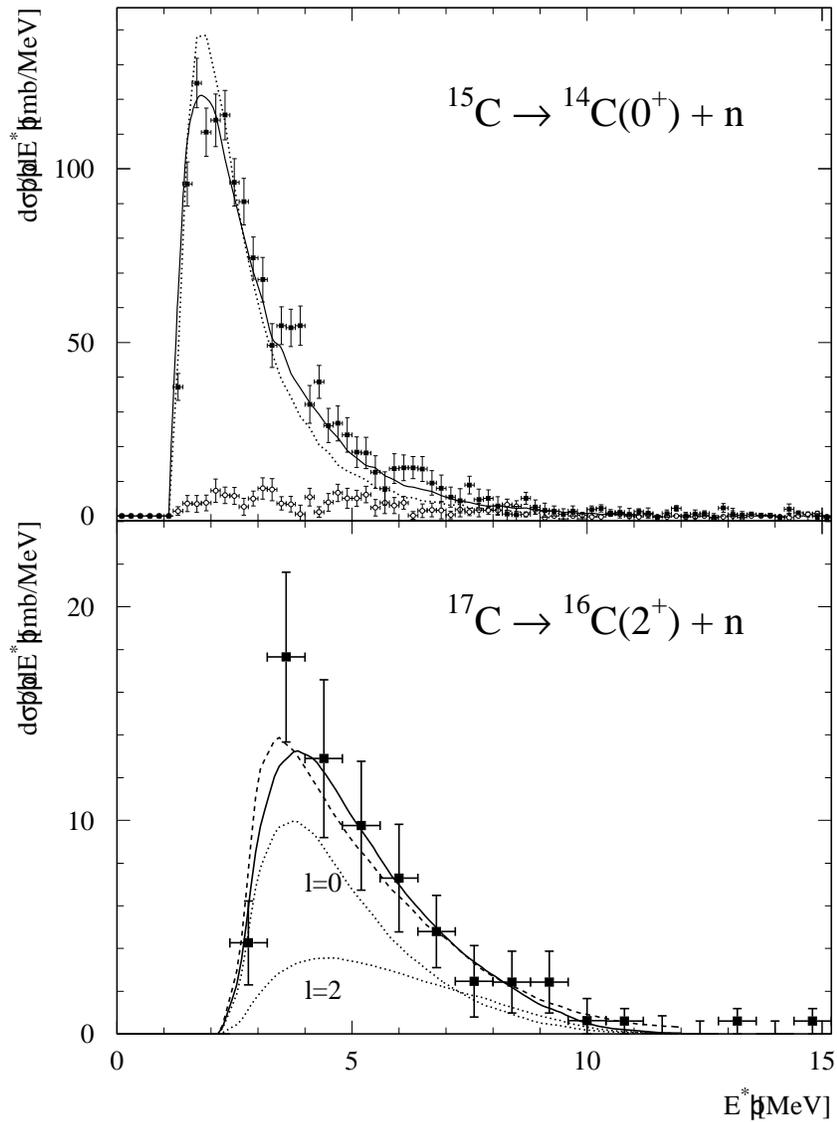}
\end{center}
\caption{\label{fig:c15c17}Differential cross sections for Coulomb
dissociation with respect to excitation energy, $E^*$ of $^{15}$C
and $^{17}$C on a lead target at beam energies of 605 and 496
MeV/nucleon, respectively \cite{datta}. See text for details of the
different curves.}
\end{figure}

Another way of studying the single particle structure of exotic
nuclei is via Coulomb dissociation at fragmentation energies of
several hundred MeV per nucleon. Using the same technique as
developed by the MSU group on knockout reactions, experiments at GSI
also measure the decay neutron and any $\gamma$-ray in coincidence
with the projectile. The differential Coulomb dissociation cross
section, $d\sigma/dE^*$, is then the incoherent sum of components
corresponding to the different core states populated following the
removal of the neutron. Figure (\ref{fig:c15c17}) shows such cross
sections for the dissociation of both $^{15}$C and $^{17}$C into
$^{14}$C and $^{15}$C, respectively. In both graphs, the solid
curves are the result of a direct breakup model using a plane wave
approximation \cite{datta}. The dotted curve in the upper graph and
the dashed curve in the lower graph both correspond to a distorted
wave approximation analysis.

Many of the techniques to describe one or two nucleon knockout
reactions as a spectroscopic tool for studying dripline nuclei are
still under development. For example, when the surviving fragment is
a halo state, whereas it was more tightly bound within the
projectile (prior to the knockout of a valence neutron), it is
necessary to include an overlap (or `mismatch') factor due to the
change induced in the remaining valence neutrons binding energy.
Since this field is relatively new, it is not appropriate to discuss
it further in this review. However a recent review of the work in
this area can be found in ref.(\cite{hantos}).

\section{Transfer reactions}
\label{transfer}

The history of transfer reactions with radioactive beams for
studying the structure of exotic nuclei is a relatively recent one.
Thus there have not been many specific developments of transfer
theories with the aim of dealing with halo-like nuclei. We will here
present the few applications that have been performed with
conventional methods and discuss the additional developments that
have been proposed.

\subsection{DWBA}

Traditionally, DWBA (Distorted Wave Born Approximation) has proved
to be extremely useful to extract spectroscopic information in
Nuclear Physics, thus one can find it in most text books
\cite{satchler}. As it retains only the first term of the Born
series, the transfer process is performed in one step.  One needs to
determine the initial and final distorted waves describing the
relative motion of projectile and target. The intrinsic structure
information is contained in the spectroscopic factor, and multiplies
the full DWBA cross section. This factor can then be related to the
structure information calculated in microscopic models \cite{brown}.

In DWBA, the transition matrix element for the transfer reaction
$A(X,Y)B$, where $A=B+v$ and $Y=X+v$,
can be written in post form,
\begin{equation}
M^{post} = < \Psi_f^{-} \; I_{YX}\;|\; V_{vB}\; +\; V_{BX}-U_{BY}\;
| \; I_{AB} \; \Psi_i^+>,
\end{equation}
and in prior-form
\begin{equation}
M^{prior} = < \Psi_f^{-} \; I_{YX}\; |\; V_{vX} \; + \; V_{BX} -
U_{AX} \; | \; I_{AB} \; \Psi_i^+>.
\end{equation}
Here, $I_{AB} = \phi(r_{v-B})$ is the overlap function of the
composite nucleus $A$ and its core $B$, and  $I_{YX}= \phi(r_{v-X})$
is the overlap function of the composite nucleus $Y$ and its core
$X$. The distorted waves $\Psi_f^{-}$, and $\Psi_f^{+}$ are
calculated using the corresponding optical potentials $U_{BY}$ and
$U_{AX}$. Often the remnant term in the transition operator
($U_{rem}^{post}=V_{BX} - U_{BY}$ or $U_{rem}^{prior}=V_{BX} -
U_{AX}$) is neglected. Furthermore, the zero range approximation
may be applicable to either $V_{vB}$ or $ V_{vX}$.

Even when applied to reactions with stable beams, the DWBA was
considered to have a limited accuracy ($\approx 30$\%). One of the
reasons for this restriction is due to uncertainties in the optical
potentials responsible for distorting the incoming and outgoing
waves. Typically, one makes use of elastic scattering data taken
over a wide angular range, to pin down the optical parameters. With
radioactive beams, these data are not available, and in some cases
even not measurable. So far, in most applications to exotic nuclei,
optical potentials have been obtained via several methods: either
extrapolated from nuclei in the valley of stability (global
parametrisations), or from double folding models involving
projectile and target densities, or have been determined with
elastic data taken only at forward angles, where the sensitivity to
the parameters is low \cite{trache}. As a consequence, these optical
potentials may bring about large uncertainties \cite{jcf}.

In addition to the distorted waves, one also needs to worry about
the transfer transition operator. So far results show that zero
range DWBA should not be used for halo nuclei. Also, the core-core
interaction for these dripline nuclei may differ significantly from
the potential describing the scattering of the unstable nucleus from
the target. Then there is no cancellation of the remnant term in the
transfer operator. For loosely bound systems, finite range effects,
as well as the remnant contribution, have been shown to be important
\cite{jcf}.

Often the cores of dripline nuclei are stable and
can be used as targets. In those circumstances stripping reactions
can populate states of the exotic nucleus, thus providing
some spectroscopic information. However, in order to get
the full spectroscopy of the ground state
the exotic nucleus should be used as the beam.

The initial attempt to extract the structure of an exotic nucleus
using a transfer reaction was for $^{11}$Be through a (p,d) reaction
in inverse kinematics \cite{winfield}. Being the first of what we
expect will be a series of experiments, it is important to
understand the approximations performed in the calculations used for
the analysis of the data, and identify the accuracy of the approach.

First, let us review briefly what is known about the ground state
structure of $^{11}$Be. It is well accepted that the loosely bound
neutron is mainly in an $s_{1/2}$ state, but there is also a
significant core excited $^{10}$Be($2^+$) component, where the
neutron is found in a $d_{5/2}$ state. In \cite{winfield} these
components are initially calculated separately, and added
incoherently, in the Separation Energy prescription. Keeping in mind
that the deformation of the core is very strong in this system,
dynamical effects are bound to play a role. In such cases, the
components for the ground state wavefunction should be calculated
properly, in a coupled channel description.

Secondly, the choice of optical potentials needs to be considered.
The proton optical parameters are taken from proton elastic
scattering data of stable nuclei, and there is no evidence that this
is a correct procedure. Although the deuteron-$^{10}$Be elastic
scattering has been described at nearby energies with the optical
model, there is the possibility of the deuteron breaking up in the
process. Then the ADBA potential (Adiabatic Deuteron Breakup
Approximation) may be more adequate \cite{adia}. Note that adiabatic
models were discussed in section \ref{adiamodel}. The difference
between these two potential choices is considerable. Unfortunately,
the transfer cross sections are very sensitive to the parameters of
both entrance and exit potentials. It seems that only additional
elastic and breakup data, taken exactly at the relevant energies,
would reduce all these uncertainties.

Similar concerns on the possible inadequacy of the optical
potentials could be expressed for the extraction of
spectroscopic factors for $^{17}$F, from the analysis
of $^{16}$O(d,n) data \cite{lewis}.

The ANC method offers an indirect measurement for the low energy
capture rates needed in astrophysics, through transfer reactions.
DWBA is widely used in the extraction of ANCs (Asymptotic
Normalisation Coefficients) \cite{anc}. The essential condition for
the applicability of the ANC method is that the reaction should be
completely peripheral, so it will only probe the asymptotic part of
the overlapping wavefunctions. The applications have concentrated on
$^8$B \cite{b8anc}, although there are many ongoing projects to
measure ANCs for other loosely bound nuclei. Essentially, there have
been two independent sets of measurements: those on the very light
targets, i.e. $(d,n)$ and ($^3$He,d), or those on heavier targets
(typically the stable Boron to Oxygen isotopes). In $(d,n)$, even if
the reaction is peripheral, the transfer cross section depends very
strongly on the choice for the optical potentials, and typically
elastic scattering corresponding to the exit channel cannot help in
pinning down the parameters \cite{jcf}. In addition, deuteron
breakup may need to be considered. For heavier targets, the
dependence on the optical parameters is not so strong, but there are
many open channels accessible to the reaction path. Other tests of
the validity of the DWBA approach should then be performed. This
discussion will continue in the following section.

For three body projectiles, the partial
wave decomposition involves the coupling of four angular momenta
and a converged calculation can easily become unfeasible.
In \cite{ogane}, the partial wave decomposition is avoided by
performing the nine-dimensional integral corresponding
to the DWBA transition amplitude \cite{ogane}. Results for
two nucleon transfer of 151 MeV $^6$He on proton and alpha particles
are extremely promising. By including the three-body structure into
the reaction calculation, this work shows how
the details of the halo ground state are determinant in the
reaction process.

\subsection{Coupled channels}

There are many ways of going beyond the one-step DWBA approach,
and we shall mention a few here.

If there are strongly coupled open inelastic channels in the
entrance (exit) partition, one can still treat the transfer process
in first order, but allow for various steps between the relevant
entrance (exit) channels. This provides an n-step DWBA method which
becomes the CCBA method (Coupled Channel Born Approximation) when
the inelastic couplings are treated to all orders.

\begin{figure}
\begin{center}\includegraphics[%
	width=0.8\columnwidth]{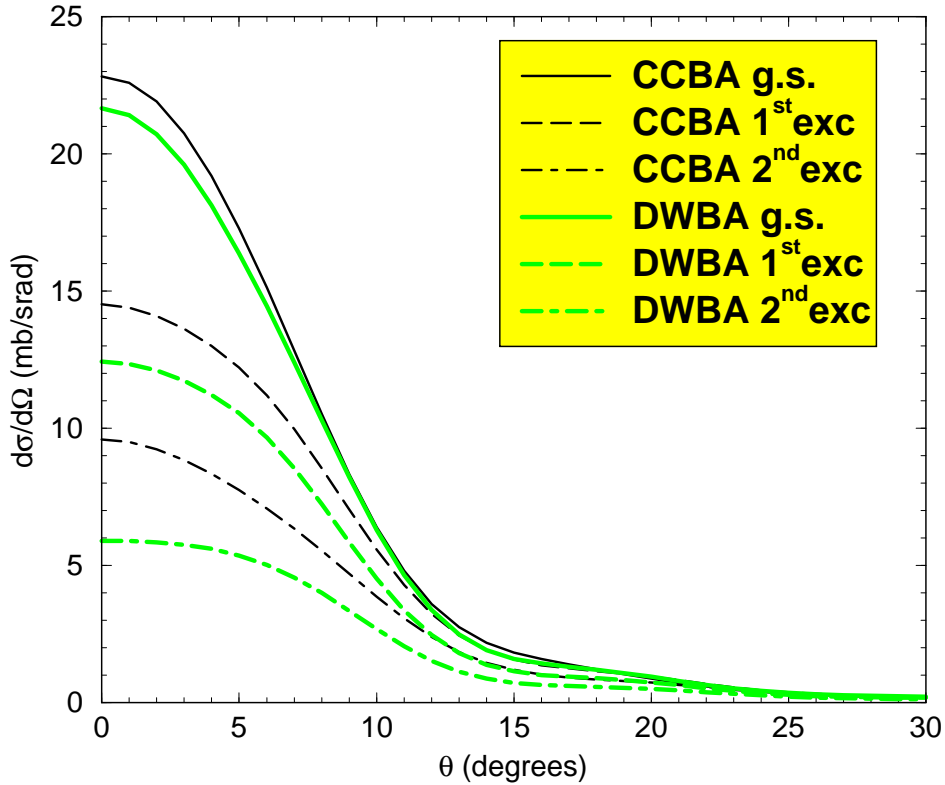}
\end{center}
\caption{\label{fig:anc}The transfer cross section for
$^{10}$B($^7$Be,$^{8}$B)$^9$Be at 84 MeV: grey lines are
the result of DWBA calculations
whereas black lines are those for CCBA. }
\end{figure}
Intuitively one would guess that transfers on well deformed targets
require a CCBA reaction model instead of the DWBA, as inelastic
couplings are known to be strong. This was confirmed by the coupled
channel tests performed in \cite{anc-cc}. In figure (\ref{fig:anc})
the transfer to the three first states in $^9$Be are predicted. The
differences between the grey lines and the black lines gives an
estimate of the error made when using the one-step approach.

Alternatively, one can also have couplings within the radioactive
projectile, due to inelasticities of the core. There should
obviously be consistency in the structure and the reaction models
used. If the structure model predicts, as in the case of $^{11}$Be,
a significant core excited component, then core excitation needs to
be included in the reaction model through an n-step DWBA or
preferably a CCBA formalism. The large two-step DWBA contribution to
the $^{11}$Be$(p,d)$ reaction estimated in \cite{winfield} does not
come as a surprise.

One should still consider the proximity to threshold of these exotic
projectiles. This may provide strong couplings to the continuum
which in principle can affect the transfer process. As mentioned
before, breakup cross sections for these nuclei are generally large
and may feedback to the transfer cross sections. In such cases, one
of the standard theoretical approaches to handle the problem is the
so-called CDCC-BA method,  where the continuum is appropriately
discretised and fully coupled,  but the transfer process is still
treated in first order Born approximation. CDCC-BA calculations were
performed for the reaction $^{14}$N$(^7$Be$,^8$B$)^{13}$C
\cite{anc-cdcc,b8br} and the results show that for this system, the
transfer is not affected by couplings to the continuum, in
particular in the forward angle region.

The complexity of the problem increases when there is the
possibility of breakup in both entrance and exit partitions, such as
is the case for $^7$Be$(d,n)^8$B. The first attempt of including
both deuteron and $^8$B breakup in this reaction was presented
recently \cite{ogata} although many approximations were involved in
the simplified version of the CDCC-BA-CDCC model. In these
calculations, the basis is over complete, and orthogonality issues
need to be carefully considered. Also, such a calculation sits at
the limit of computational capacities. In spite of this, more
calculations along these lines will be needed in the future.

If the transfer is very strong, then first order Born series may not
be sufficient. One can then allow for multi-step transfer through
the CRC (coupled reaction channels) method. Applications to transfer
reactions with $^{11}$Be \cite{winfield} and $^8$B \cite{jcf} show
that such higher order multi-step terms provide less than $10$\%
corrections.

\begin{figure}
\begin{center}\includegraphics[%
	width=0.8\columnwidth]{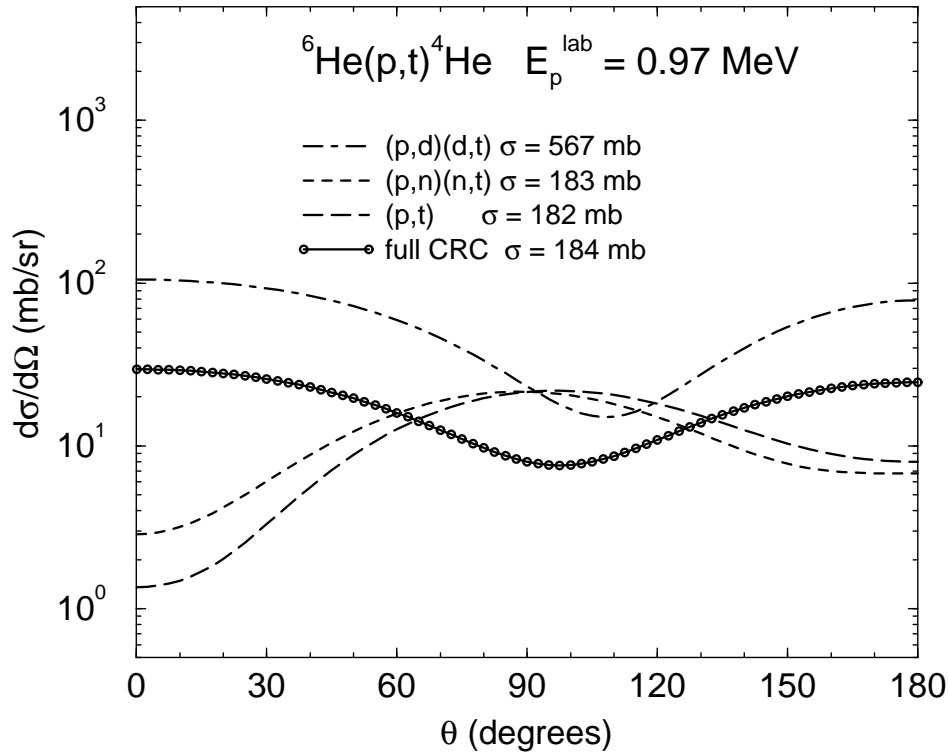}
\end{center}
\caption{\label{fig:he6pt}The 2-nucleon transfer of $^{6}$He
on protons.}
\end{figure}
When more than one nucleon is transferred, one-step transfer may be
supplemented by two- and higher order step-processes. When the
reaction time is large, the system can rearrange itself in several
ways. Inevitably one can find many transfer paths for the reaction,
which need to be coupled in a CRC method. The low energy
experimental program using the  $^6$He beam in LLN \cite{he6lln}
motivated the application to $^6$He$+p$ \cite{timo2}. In figure
(\ref{fig:he6pt}) the various multistep paths relevant to the
reaction are shown as well as the total cross section, resulting
from the interference of the considered channels. Calculations were
performed within a CRC formalism. In those calculations, one and
two-nucleon transfer form-factors were determined within a
three-body structure model and a full finite-range treatment was
included. Both the remnant term and nonorthogonality corrections
were found to be necessary. Couplings to all open
channels were important to generate the final cross sections.

\subsection{Other methods}

When the transfer process occurs at sufficiently high energy, a
Glauber approach is possible \cite{carstoiu}. The sensitivity to
details of the projectile wavefunction is shown to be very strong in
the calculations for $^{11}$Be$(d,p)^{10}$Be, although only a single
particle form factor is assumed. In fact, one expects that the
sensitivity is more on the scale than the shape details and
components. It would be interesting to confirm this by improving the
structure information included in the reaction model.

\begin{figure}
\begin{center}\includegraphics[%
	width=0.8\columnwidth]{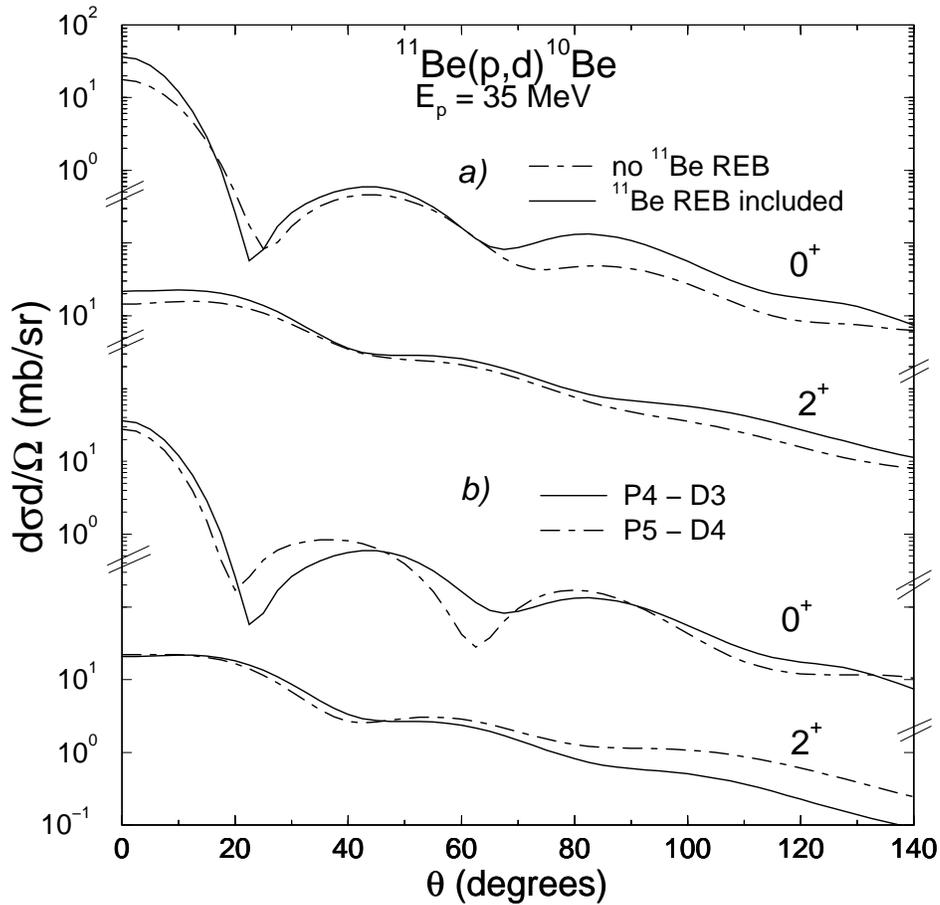}
\end{center}
\caption{\label{fig:natasha}The transfer reaction
$^{11}$Be(p,d)$^{10}$Be$(0^+,2^+)$:
a) comparing the effect of REB for a fixed set of optical potentials;
b) comparing two choices of optical potentials while including
the REB effect. In both cases the deuteron breakup was included.}
\end{figure}
If, on the other hand, the energy is not very high, but still high
enough compared to the binding energy, an adiabatic approximation
can be safely made. An adiabatic model was developed for one-nucleon
halo nuclei \cite{timo1}. The T-matrix is written so that only the
halo-nucleon/target interaction appears in the operator. Then both
core-halo and target-core interactions need to be considered in
calculating the entrance(exit) wavefunction. The adiabatic solution
is a distorted wave that includes recoil and breakup effects (REB).
The exact wavefunction appearing in the exit(entrance) channel is
approximated to a distorted wave and the adiabatic limit is also
taken (making use of the Johnson-Soper potential \cite{adia}).
Calculations for $^{16}$O$(d,p)^{17}$F, $^{10}$Be$(d,p)^{11}$Be and
$^{11}$Be(p,d)$^{10}$Be are presented. The REB effects generally
increase the cross section, so that the extracted spectroscopic
factors are generally reduced. In all cases the impact of deuteron
breakup on the transfer cross section was stronger than that of the
heavier nucleus. The example for $^{11}$Be(p,d) populating both the
g.s. and the first $2^+$ in $^{10}$Be is shown in figure
(\ref{fig:natasha}). Disappointingly, the spectroscopic factors
extracted for $^{11}$Be g.s. do not agree with those in
\cite{winfield}, although a direct comparison with this data was not
performed. It would be interesting to include core excitation in
this reaction model, as the d-wave dynamics are expected to change
the picture. Also, a comparison with the various standard models is
necessary to better understand the advantages and drawbacks.
Nevertheless this model is very promising, since in principle it can
deal with breakup in both entrance and exit partition and yet is not
computationally demanding.

Sometimes, transfer methods can be applied to breakup reactions. The
usual way to think about projectile breakup ($P\Rightarrow C+x$) is
as $T(P,C+x)T$. However, the transfer to the continuum of the target
$T(P,C)T+x$ is formally equivalent to this, and the calculation may
converge more quickly. If the experiments cannot distinguish the two
processes apart, then the transfer description (including bound and
unbound states) may become more attractive. This approach offers the
best description to date of the $^6$He 2n low energy
transfer/breakup on $^{209}$Bi \cite{he6nd}.  It is also proving to
be very promising for the $^8$Li transfer/breakup on $^{208}$Pb
\cite{li8moro}.

\section{Fusion}
\label{fusion}

Theoretical developments for fusion reactions, specifically designed
for light dripline nuclei, are still scarce, partly due to the fact
that accurate fusion data on these nuclei is rather recent, and
partly because fusion makes up such a small fraction of the reaction
cross section. For this reason we first point out a few general
aspects concerning the theory of fusion reactions with stable
nuclei, of relevance for loosely bound systems, and discuss the
theoretical advances when applied to RNBs.

\subsection{Some relevant ideas from Heavy-ion fusion}

It was only in the early eighties that the heavy-ion fusion data
allowed the refinement of fusion calculations, which now go much
beyond the basic barrier penetration ideas \cite{wong}, and
incorporate coupled channel effects of various types (e.g.
\cite{broglia,dasso1,dasso2,landowne,lindsay,ian-fus}). There are
essentially two approaches: a) the fusion process  is modeled with a
strong imaginary potential in the interior, taking into account the
loss of flux from all other channels and b)  the incoming boundary
condition method, where each component of the wavefunction is
matched to an incoming wave on the barrier. In both cases a coupled
channel equation is solved. The first approach generally contains
very strong imaginary potentials which remove any couplings acting in
the barrier region. Under these circumstances, as couplings are
limited to larger distances where they are typically weaker, the
DWBA first order solution may be adequate to describe the mechanism.
However, in the strong coupling limit, the determination of the
imaginary potential is by no means transparent and coupling effects
may be misinterpreted \cite{dasso1}. It is often preferrable to use
the incoming boundary condition, both for technical and physical
reasons. Often one also takes the adiabatic approximation, when
excitation energies of the colliding nuclei are negligible compared
to the fusion energy, and the differences in the centrifugal
barriers can be ignored \cite{lindsay}. Lindsay {\it et al.}
\cite{lindsay} showed that, in the strong coupling limit, the
adiabatic calculations reproduce the full coupled channel results,
while the DWBA calculations overestimate the fusion cross section.
Explicit simplified expressions in terms of the one channel cross
section have been deduced for vibrational and rotational structures
\cite{lindsay}.

It is well understood that in general the inclusion of
extra channels, coupled into the reaction mechanism,
produce more steps in the barrier distribution and consequently
spread the energy range over which the transmission factor
goes from zero to unity. This produces enhancement of sub-barrier
fusion, and hinders fusion above the barrier \cite{dasso1,lindsay}.

Also, while the flux transmitted by the barrier in a particular
channel (fusion) depends on the strength of the interaction through
the barrier, the reflected flux, the reaction cross section that is
experimentally seen, depends on the couplings outside the barrier.
This means that the connection between the fusion part and reaction
part of the flux for each channel is not straightforward.

It has been shown that the energy matching for the channel to
be coupled, relative to the incident channel, is a crucial ingredient.
In fact, negative Q-values tend to reduce the relative enhancement of
sub-barrier fusion (when comparing with the perfectly matched
channels Q=0), and positive Q-values tend to enlarge these effects
\cite{dasso1,dasso2}. However, negative Q-values produce
an increase of the overall fusion flux and positive Q-values
an overall decrease, when compared with $Q=0$.

The inclusion of inelastic couplings alone is frequently not enough
to describe the data \cite{landowne, timmers} and transfer couplings
have been suggested as the necessary solution.  Neutron transfer has
often been a successful explanation for the enhancement of
sub-barrier fusion cross sections (see for instance the results on
the fusion of $^{58}$Ni+$^{64}$Ni, relative to $^{58}$Ni+$^{58}$Ni
\cite{broglia}). One can expect that for loosely bound neutron rich
(or proton rich) nuclei this effect will become even more important.

Given the influence of both inelastic couplings and
transfer couplings, fusion calculations can easily become
rather demanding. The CRC method (coupled reaction channels)
mentioned in section \ref{bu} offers a reliable path for the fusion
calculation \cite{ian-fus}. In the end, a consistent
description of the elastic, inelastic and transfer channels
should be obtained, at the same time as the fusion
cross section.

The number of data sets on the fusion of light dripline nuclei is
increasing by the day, yet we are still far from understanding the
general behaviour \cite{alamanos}. Given the importance of breakup
and transfer channels, experiments designed to measure specifically
these components have been carried out (e.g. \cite{he6nd}).
Nevertheless, many studies are still presently performed on the
stable Li  or Be isotopes, where systematic trends can be more
easily identified \cite{takahashi}. Below we discuss the theoretical
contributions in this field, starting with the simple models
initially used, up to the full CDCC or the time-dependent models,
already applied in earlier sections.

\subsection{Preliminary fusion results}

Interest in the fusion of dripline nuclei was initiated
in the early nineties for two reasons: to better understand
the exotic properties of these nuclei and for the hope
of insight into the production of the superheavy
elements. The first sequence of theoretical work was applied
to $^{11}$Li \cite{hussein1,dasso3, hussein2} for which
there are as yet no available data.

In \cite{hussein1}, a two-channel fusion calculation
included a resonance at around 1.2 MeV
(referred to as a pygmy resonance, or giant dipole resonance).
The possibility of breakup was accounted for through a survival
probability multiplying factor. The idea
was that breakup channels take flux away from fusion.
Even though rather poor, the adiabatic approximation was used.
The conclusion was that fusion is suppressed above the barrier,
and the prediction of sub-barrier enhancement is lower than what
would be obtained if no breakup was included. It was subsequently
pointed out that the breakup channel need not reduce fusion
\cite{dasso3}. In fact, the additional coupling could produce
enhancement in the same way inelastic or transfer couplings do.
The coupled channel calculations in \cite{dasso3} determine
the complete fusion of $^{11}$Li on $^{208}$Pb for the
same barrier parameters as those used in \cite{hussein1}. Results
show enhancement due to the dipole coupling and further enhancement
due to the coupling to breakup states. A theoretical
improvement of the treatment of resonant channels in
\cite{hussein1} uses doorway states \cite{hussein2}. The results
confirmed fusion hindrance around the barrier,
sustaining the controversy.

As knowledge of the properties of light exotic nuclei increased,
breakup and transfer channels became an unavoidable issue when
considering the fusion process. Many of these nuclei do not have
excited states and the inclusion of inelastic excitation of the
target is  by now a standard procedure. However, given the loosely
bound nature of these beams, the coupling strength to continuum
channels is expected to be strong and the Q-value for neutron
transfer (for neutron rich) or proton transfer (for proton rich) can
become positive, which if well matched can also produce strong
coupling.

In \cite{oertzen}, a study of the effect of transfer and inelastic
couplings on the fusion of $^{11}$Be$+^{12}$C is performed. The CRC
calculations included both bound states and the d-wave resonance of
$^{11}$Be, as well as several one-neutron transfer channels to
$^{10}$Be$+^{13}$C, corresponding to the three first states in
$^{13}$C and the first two states of $^{10}$Be (note that all have
positive Q-values). The equations were solved using a strong but
short-range imaginary potential in order to remove coupling from
small distances. It was found that the transfer alone decreases the
fusion and that inelastic couplings partly compensate this
reduction. Coupling effects dissappear at higher incident energies
while the maximum effect is found just below the barrier.

\subsection{The time-dependent method}

One can also determine the fusion cross section using the time
evolution of a wavepacket, solution of the time-dependent
Schr\"odinger equation. Such calculations were performed by Yabana
\cite{yabanafusion} for a core+n system impinging on a target, that
simulates the $^{11}$Be$+^{40}$Ca reaction. Due to computational
limitations, nuclei are considered spinless, only l=0 relative
motion is included, and the wavepacket is confined to a finite
radial region where only the Coulomb part  of the core-target
interaction is felt. The absorption is included through the
imaginary part of the core-target interaction alone, thus the fusion
cross section contains both complete and incomplete fusion. The
behaviour of the process for low binding ($-0.6$ MeV) is compared
with a system with stronger binding ($-3$ MeV). Breakup, transfer,
and fusion cross sections are simultaneously calculated.

It was found in that study that, by increasing the depth of the
neutron-target potential until binding is larger than the binding of
the neutron in the projectile, the projectile-target kinetic energy
is increased, producing an increase of the fusion cross section. The
opposite effect happens when the n-target interaction produces less
binding for the n-target system than the separation energy of the
neutron in the projectile. However, when the projectile binding is
already very low, the likelihood is that the neutron will have a
larger binding to the target, resulting in an overall suppression of
fusion. Fusion cross sections as a function of beam energy show a
slight enhancement below the barrier and a clear reduction above the
barrier.

\subsection{The CDCC method}

\begin{figure}
\begin{center}\includegraphics[%
	width=0.5\columnwidth]{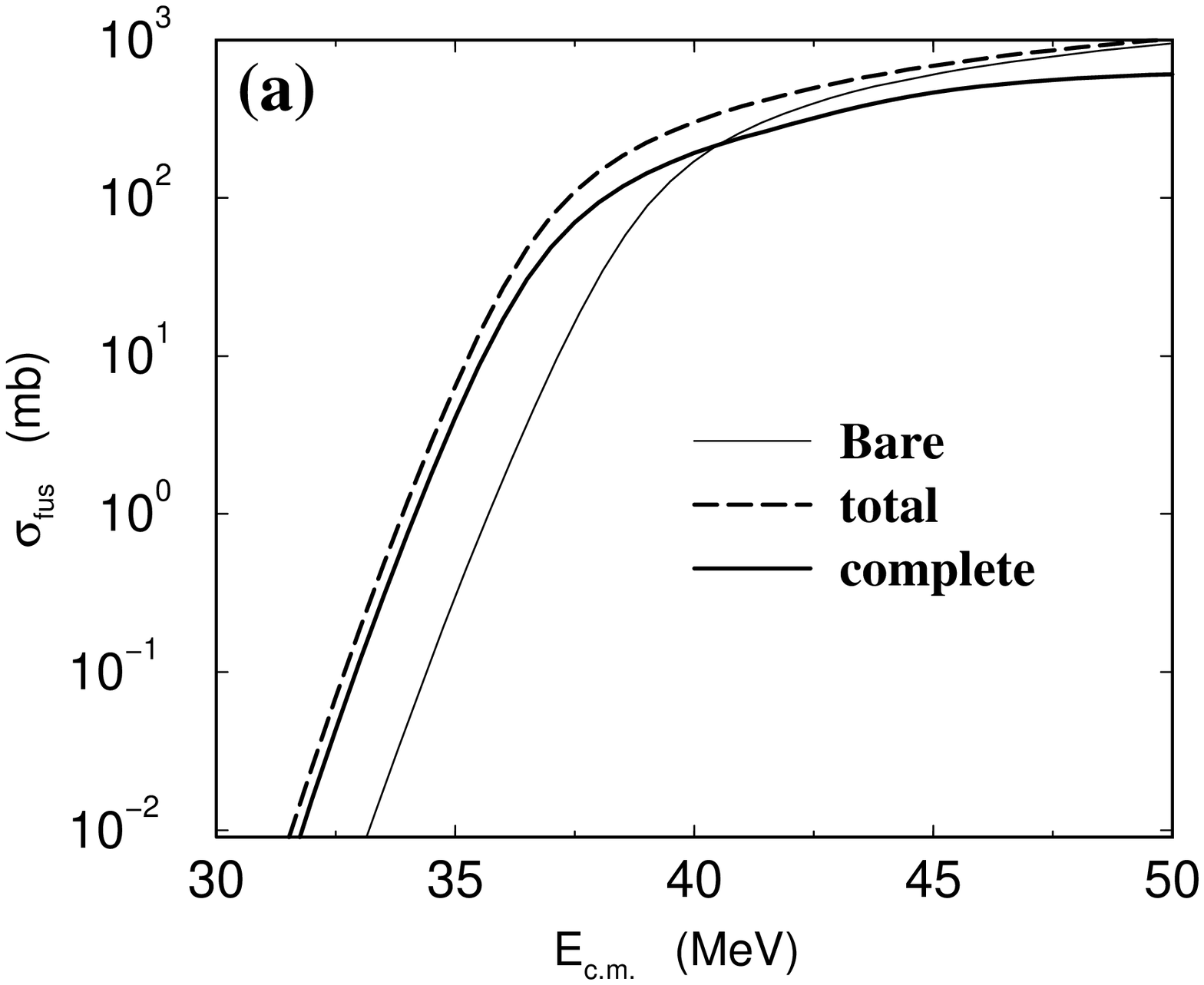}
\end{center}
\begin{center}\includegraphics[%
	width=0.5\columnwidth]{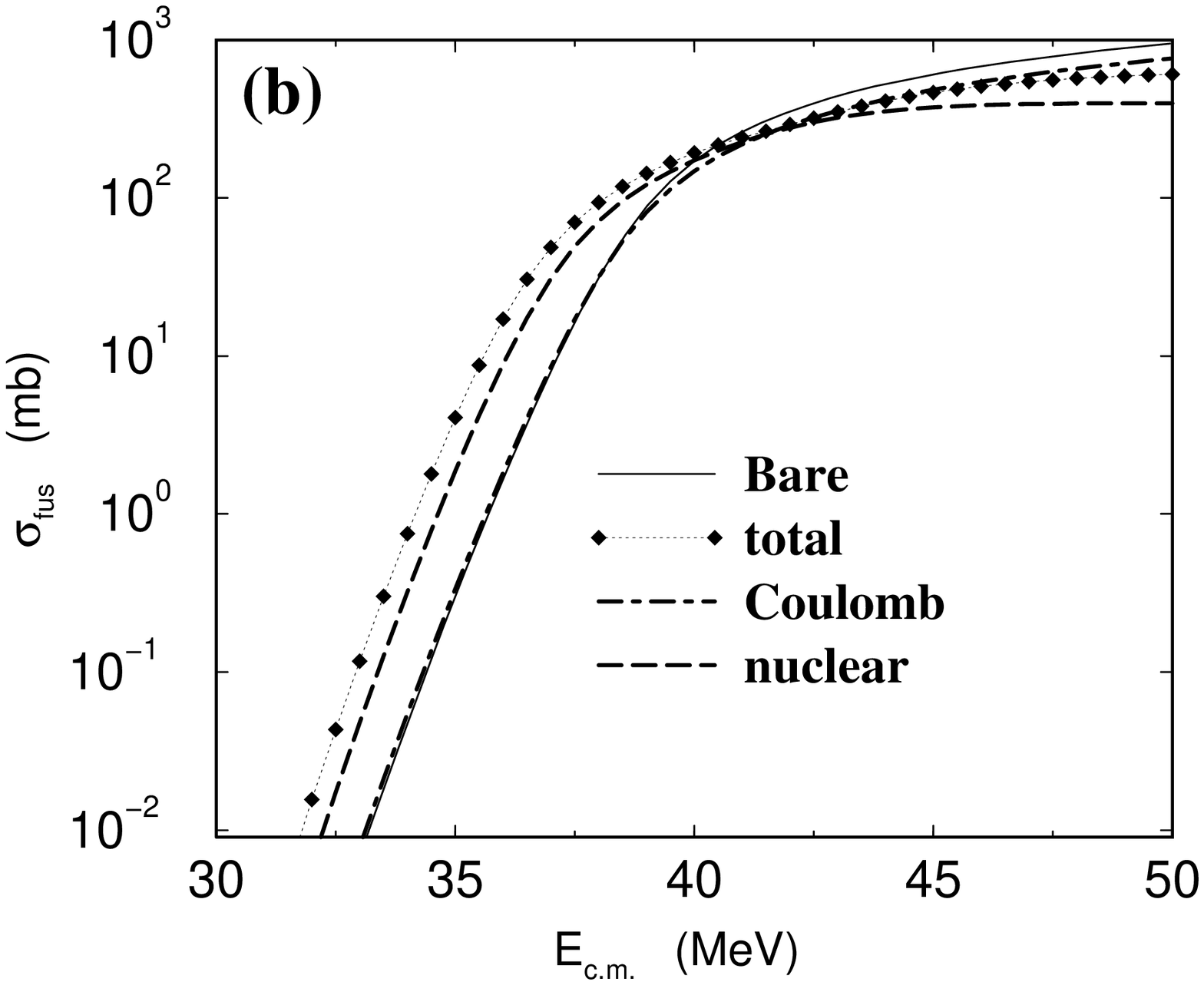}
\end{center}
\caption{\label{fig:hagino}The fusion cross section for
$^{11}$Be+$^{208}$Pb as a function of bombarding c.m. energy: (a)
complete fusion (thick solid) and the total fusion (dashed) are
compared with the simple penetration model (thin solid); (b) the
nuclear (long-dashed) and Coulomb (dot-dashed) contributions to the
complete fusion (diamonds). The barrier penetration model is shown
(thin solid) for comparison.}
\end{figure}

The CDCC method is another possibility for including breakup effects
when calculating the fusion reaction. In the CDCC calculations for
the fusion of $^{11}$Be and $^{208}$Pb performed by Hagino {\it et
al.} \cite{hagino1}, several truncations are made in order to
concentrate on pure breakup effects (results are shown in figure
\ref{fig:hagino}). Continuum-continuum couplings are ignored, as
well as projectile and target inelastic excitations. The neutron in
the projectile is assumed to be in a pure $2s_{1/2}$ state and only
transitions to $p_{3/2}$ are considered. The incoming boundary
condition method is used for solving the coupled channel equations,
and the isocentrifugal approximation is made. The results predict
fusion enhancement below the barrier and suppression above the
barrier, in agreement with the early results of \cite{dasso1}. The
dynamical effects of the couplings are essential. It is shown that
the coupling form factors at around barrier distances, peak at
relatively large energies, behaving in a completely different way
than the asymptotic coupling form factor. The same conclusions were
obtained for the fusion of $^{6}$He and $^{238}$U for an identical
calculation \cite{hagino2}.

More recently a full CDCC calculation, without the above mentioned
truncations, was performed in order to calculate the fusion of
$^{11}$Be and $^{208}$Pb \cite{diaz}.  Although inelastic
excitations of the target and transfer couplings were left aside,
the calculation included the projectile excited state $1/2^-$ as
well as all multipoles in the continuum needed for convergence. The
fusion cross sections were defined in terms of a short-range
imaginary potential. It was found that the excited $1/2^-$ state of
$^{11}$Be has little influence on the fusion, redistributing the
fusion cross section. If no continuum-continuum couplings were
included, the conclusions of \cite{hagino1} for complete fusion were
corroborated: strong enhancement below the barrier and hindrance
above the barrier. However, since incomplete fusion is relatively
large above the barrier, the total fusion cross section, for this
energy region, was not reduced, when compared with the no-couplings
case. The truly surprising result was the effect of
continuum-continuum couplings: the reduction of the complete fusion
cross section was of nearly two orders of magnitude in the
sub-barrier region and around an order of magnitude above the
barrier. In addition incomplete fusion is also reduced. This means
that the sub-barrier enhancement is much weaker, for the complete
and total fusion processes and there is suppression above the
barrier for both complete and total fusion.

Note that, in \cite{diaz}, complete fusion is associated with
absorption from the bound channels only, when in principle it is
possible that the projectile suffers complete fusion even after
breaking up. Also, incomplete fusion is associated with absorption
from the breakup channels, and so the real incomplete fusion could
be lower than that calculated in \cite{diaz}. Nonetheless, the total
fusion cross section is unambiguous and can be compared with
experiment. Agreement is found below the barrier but cross sections
are underpredicted above the barrier.

A better understanding of the role of the continuum-continuum
couplings is probably necessary to be able to improve the
theoretical description. In addition, the separation of complete
fusion from incomplete fusion in both the data and calculations
would also be helpful. Given the conclusions in \cite{oertzen},
transfer channels will inevitable need to be included in a CDCC-CRC
type calculation before a definite conclusion can be drawn, albeit
the calculations presented in \cite{diaz} were already at the limit
of our best computational possibilities. At present, the extraction
of structure details from fusion data seems to be unlikely.
Notwithstanding, it still offers one of the best alternatives to
learn more about the production of superheavy elements.

\section{Charge exchange and photonuclear reactions}
\label{other}

Another way to study the structure (both bound state and continuum)
of exotic nuclei is through charge exchange reactions. However,
where measurements have been made for $(n,p)$ reactions, for
instance $^6$Li$(n,p)^6$He \cite{brady} and $^{14}$Be$(n,p)^{14}$B
\cite{takeuchi}, the experiments have suffered from poor statistics
and energy resolution such that individual final states could not be
resolved.  Alternatives to this are the $(t,^3$He$)$ and
$(^7$Li,$^7$Be$)$ reactions. The reaction $^6$Li$(t,^3$He$)^6$He has
been studied at MSU \cite{nakamura} and the continuum structure of
$^6$He probed. However, little theoretical work has been carried
out. Ershov and collaborators \cite{ershov2} have studied the
$^6$Li$(n,p)^6$He within a four-body DWBA model and also point out
that $^6$He$(p,p')^6$He is a useful complimentary probe. A few-body
eikonal model for charge exchange has been developed by the Surrey
group but has so far been applied only to the $(d,2p)$ reaction
\cite{rugmaice}.

In recent years, there has been a growing interest in the use of
charged pion photoproduction reactions, $(\gamma,\pi^\pm)$, as a
tool for studying nuclear structure. In particular, because of
recent advances in intermediate energy ``electron'' factories, and
the development of high resolution pion spectrometers, precise
angular distributions for the produced pions can be measured and, it
is anticipated, individual final states of the nucleus of interest
can be resolved. As discussed in this review, most of what is known
about exotic nuclei has been obtained through fragmentation
reactions in which the strong interaction, particularly that of the
core, plays a major role. Pion photoproduction studies offer an
alternative, cleaner, electromagnetic probe of nuclear structure. In
such charge exchanging reactions (e.g
$^6$Li$(\gamma,\pi^+)^6$He) the pion energy and momentum can
provide information about the valence nucleon participating in the
reaction $\gamma N \rightarrow \pi N$.

A major advantage of such reactions is that they can probe the
entire nuclear wavefunction, unlike the surface-dominated
fragmentation reactions which tend to only probe the wavefunction
tail. Of course, depending on the momentum transferred in the
process, such reactions can also probe the surface and be used to
study the halo, without ambiguities due to any core potential.
Karataglidis {\it et al.} \cite{karpion} have carried out a DWBA
calculation in which the nuclear transition density is obtained
using the shell model, in a similar way to the group's studies of
proton elastic and inelastic scattering \cite{Karatpp}.

Several years ago it was suggested \cite{katandben} that excited
state halos can also be probed in this way. It was shown that the
pion cross sections calculated for the
$^{17}$O$(\gamma,\pi^-)^{17}$F($\frac{1}{2}^+$,0.495 MeV) reaction
is sensitive to the halo structure of the valence proton.

Finally, an alternative photonuclear probe that does not involve
charge exchange is via the $(\gamma,p)$ reaction. Boland {\it et
al.} \cite{boland} have observed a broad resonance in the $^6$He
continuum at 5 MeV but are unable to define its exact nature. Such
reactions are in need of further theoretical analysis as well as
more accurate measurements.

\section{Outlook}
\label{outlook}

In this review we have attempted to cover many theoretical aspects
concerning reactions with light dripline nuclei, paying particular
attention to the interplay between the structure input and the
reaction model. In order to obtain a successful description of the
reaction, specific features associated with the exotic nature of
these nuclei need to be included. Furthermore, as data becomes more
detailed and accurate, better models are required. Better models
typically imply that calculations become larger and new numerical
methods need to be developed in order to ensure progress in the
field. Given some discrepancies found between theory and experiment,
the suggestion that core excitation can dynamically interfere in the
process needs to be checked. Improvements to the single particle
models, extensively used to date, are then necessary to assess
the role of core excitation in the reaction mechanism.
Finally we comment on possible physics with electron beams.

\subsection{Continuum couplings and computation}

Continuum coupling is crucial for understanding certain reaction
processes involving light exotic nuclei. These coupling effects are
best seen in breakup processes, but it has also been shown that
there is an important influence on elastic scattering and fusion
reactions. In fact, the results on fusion are so large that they
call for further investigations. Couplings to the continuum may be
less important for transfer reactions and certainly more examples
need to be analysed before any general statement can be made.

When including continuum coupling in a reaction model, one
should definitely take care of the non-resonant continuum as well as
resonant states. Couplings between two continuum states may be
equally (or even more) important as couplings between the ground
state and the continuum. Many results discussed here use the well
established CDCC method to discretize the continuum. However, given
the computational demand of the traditional CDCC calculations,
new methods are being developed. One of the most
promising alternative methods for discretizing the continuum uses
transformed harmonic oscillators (THO). Benchmark calculations
comparing the CDCC and the THO methods for the elastic scattering
and breakup of deuterons on $^{208}$Pb are very encouraging
\cite{tho}. We expect that, in the future, the optimization of the
continuum discretization will make it feasible to tackle reactions
involving three body systems, such as $^{11}$Li, by including the
three body continuum properly.

On the same lines, the very recent work presented in \cite{pseudo}
proposes a pseudo-state discretization using real and complex-range
Gaussian bases to calculate the breakup within the 3-body
CDCC picture. Applications to the 4-body CDCC problem are also under
consideration.

\subsection{Core excitation}

In recent years, the description of light exotic nuclei based
on single particle models has become unsatisfying. Although
extremely attractive for their simplicity, one needs to go beyond
the inert core approximation in order to account for the physics
that can now be accurately measured in the new facilities. We have
already mentioned the experiments involving the Be isotopes, where
excited core components were clearly identified (e.g. \cite{navin2,
winfield}). Evidence for a core excited component was also found in
the breakup of high energy $^8$B \cite{b8gsi}. Besides, similar
results are to be expected for many other nuclei.

In the light of these new experimental results, theoretical reaction
models need to encompass core excitation. As excited states of the
core involve typically larger angular momentum, there will be a
rapid increase of the number of available reaction channels. In
cases such as $^8$B where the ground state of the core is a $3/2^-$
state, the spin of the core has been routinely neglected and most
models do not allow the core structure to play any role in the
reaction mechanism. In reaction models that include core excitation,
this can no longer be done and the calculations become much larger.
Apart from the computational demands, there are some theoretical
issues that need to be addressed. Namely, since the loosely bound
systems often require the coupling to breakup states, a reaction
model with core excitation implies that both bound and unbound
states need to have core excited components. It is thus necessary to
generalize the CDCC approach to include core excitation within the
projectile.

Most efforts to include core excitation in the reaction model have
been performed in a static way. For example in
\cite{maddalena1,shyam5}, core excitation components of the initial
wavefunction are included in a DWBA calculation for the Coulomb
dissociation of the $^{11}$Be, $^{17}$C and $^{19}$C. Yet there is
no dynamical excitation of the core throughout the reaction process.
In other words, core excitation could not be generated in the
reaction. This approximation does not seem adequate, especially in
cases where the couplings to core excited states are strong. The
best attempt so far of studying the effect of core excitation in
reactions with light exotic nuclei was performed in \cite{b8br}. In
that work, core excitation in $^8$B was dynamically included in the
transfer reaction $^{14}$N($^7$Be,$^8$B)$^{13}$C. A CDCC-BA
formalism was used, employing the approximation that the continuum
bin-states could still be described within a single particle picture
\cite{b8br}. As a conclusion of that work, core excitation turned
out not to be important. One should note that there is an
inconsistency in the calculation \cite{b8br} in that the projectile
Hamiltonian for the bound state is not the same as that for the
unbound states. There are non-orthogonality issues that arise and
should be investigated. Further work on generalizing CDCC with core
excited bins is underway.

Recent results show that, as one moves away from the stability line,
few-body models are less successful. An example can be found in
\cite{be12nunes02} where many $^{12}$Be reaction observables are
compared with recent data. The three body model of $^{10}$Be+n+n in
which the core $^{10}$Be is assumed to be a perfect rotor and is
allowed to excite to the first $2^+$ state, is unable to reproduce
the correct E2 transition $2^+ \rightarrow 0^+$ for $^{12}$Be, if
all other observables are to be reproduced. This suggests that the
simple core excitation ideas need to be revisited. Microscopic
information needs to be integrated to an extent that it becomes
useful in a reaction model, and yet retains the necessary detail.
This balance is one of the most challenging problems to be dealt
with in the near future.

\subsection{New physics with electron beams}

For half a century, electron scattering experiments on nuclei have
contributed significantly to our understanding of the structure of
stable nuclei. However, since short-lived exotic species cannot form
a nuclear target that is at rest in the laboratory frame,
electron-nucleus (eA) colliders are being built \cite{gsi,riken}
which will give access to structure studies of unstable nuclei,
opening up a new era of low-energy electron scattering. Due to the
limited luminosities achieved with radioactive beams, the first
generation of experiments with these colliders will focus on
measurements of the radii of the nuclear charge distribution and its
diffuseness. Electron scattering on a large variety of unstable
nuclei will become possible and will help clarify the evolution of
charge radii towards the driplines.

Since the electromagnetic interaction is relatively weak, multiple
scattering effects can be neglected in electron scattering and the
interaction can be well described by one-photon exchange terms. By
combining the charge distributions from electron scattering with
matter distributions from hadronic probes such as proton scattering
(in inverse kinematics) it will be possible to determine proton and
neutron distributions separately for a large number of exotic
nuclei.

More interestingly, inelastic electron scattering, which  is known
to be an excellent tool for studying the spectroscopy of bound and
unbound states in nuclei, will also be possible. The transition form
factors are related to different multipolarities of the excitations
and offer a unique way of studying collective motion in unstable
nuclei.

In addition, inclusive quasielastic scattering, $A(e$,$e')$, is
well-known as a way of probing nucleon motion within the nucleus
\cite{donn}. Exclusive quasielastic scattering involves detecting a
knocked-out constituent in coincidence with the electron,
$A(e,e'x)B$. For instance, the unpolarized quasifree $(e,e'p)$
reaction has been systematically used to probe single particle
properties of complex nuclei, such as momentum distributions and
spectroscopic factors. Excellent agreement with experiment has
been obtained for these observables for both $^6$Li$(e,e'\,)^6$Li
\cite{wiringa} and $^7$Li$(e,e'p)^6$He \cite{lapikas} reactions.
The reliability of the information extracted
from such reactions is due to the weak dependence of the observables
upon the electron scattering kinematical conditions. This is
particularly true in the quasielastic region, where both the
momentum transfer and energy transfer are high enough for the
interaction between the electron and a single nucleon in the nucleus
to dominate.

Due to the novelty and relatively recent interest in this field,
there has been, as yet, very little theoretical work. The only
available calculations\cite{garr}, for electrons scattering from
$^6$He, make use of the plane wave impulse approximation PWIA (which
assumes that the virtual photon is absorbed by a single
constituent). While that study described the $^6$He as a three-body
system, the authors make several further simplifying assumptions,
such as neglecting the Coulomb distortion of the electron, and final
state interactions of the knocked out constituent with the spectator
constituents. Further work is clearly needed, and on a range of
light exotic nuclei.

\section{Summary}

The relatively new field of the structure and reactions of light
exotic and halo nuclei has certainly provided a 'wake up' call for
nuclear theorists. Textbook models have been applicable in certain
cases while elsewhere new approaches have had to be developed. We
stress though that reviews such as this act only as staging posts
along the road; there is still a challenging and no doubt
fascinating journey ahead.

\ack
The authors wish to thank Prof. I.J. Thompson for valuable comments
on the manuscript. We are grateful to T. Aumann, W. Catford, H. Esbensen,
S. Ershov, K. Hagino, P.G. Hansen, N. Timofeyuk and V. Lapoux
for kindly providing some of the figures included in the review.
The work of J.A-K is supported by the United
Kingdom Engineering and Physical Sciences Research Council (EPSRC)
through grants GR/M82141 and GR/R78633/01. Support from the
Portuguese Science and Technology Foundation under grant
POCTI/43421/FNU/2001 is also acknowledged.

\end{document}